\documentclass[preprint,preprintnumbers,amsmath,amssymb]{revtex4}

\usepackage{latexsym,amssymb,makeidx}

\usepackage{epsfig, color, ulem}
\usepackage{graphicx}
\usepackage{dcolumn}
\usepackage{bm}
\usepackage{amsmath}
\usepackage{setspace}

\def\i{i}
\def\J{j}
\def\rmd{{\rm d}}
\def\rmdd{{\rm d}}
\def\bx{{\bf x}}
\def\bxA{\bx_A}
\def\bxB{\bx_B}
\def\bxh{{\bf x}_{\rm L}}
\def\bxha{{\bf x}_{{\rm L},A}}
\def\p{P}
\def\s{S}
\def\ss{s}
\def\bq{{\bf q}}
\def\bd{{\bf d}}
\def\bp{{\bf p}}
\def\bs{{\bf s}}
\def\bN{{\bf N}}
\def\bK{{\bf K}}
\def\bJ{{\bf J}}
\def\HH{{\cal H}}
\def\bA{{{\mbox{\boldmath ${\cal A}$}}}}

\def\bL{{{\mbox{\boldmath ${\cal L}$}}}}
\def\bH{{{\mbox{\boldmath ${\cal H}$}}}}

\def\setA{\mathbb{S}}
\def\setD{\mathbb{D}}
\def\setdD{{\partial\setD}}

\def\setdDD{{\setdD_{0,1}}}
\def\setdDA{{\setdD_{0,A}}}
\def\half{\frac{1}{2}}
\def\quart{\frac{1}{4}}
\def\setDA{\setD_A}

\begin{document}
%\preprint{AMP Preprint}
\bibliographystyle{jasanum}

\def\rev#1{\textcolor{black}{#1}}

%\title{Flux-normalised versus field-normalised decomposition of the scalar wave equation}

\title{Reciprocity and representation theorems for flux- and field-normalised decomposed wave fields}
%, with applications in reflection imaging}

\author{Kees Wapenaar}
\affiliation{Department of Geoscience and Engineering, Delft University of Technology, 2600 GA Delft, The Netherlands}
\date{\today}

\begin{spacing}{1.3}
%\begin{spacing}{2}

\begin{abstract}
\noindent
We consider wave propagation problems in which there is a preferred direction of propagation. To account
for propagation in preferred directions,  the wave equation is decomposed into a set of coupled equations for waves that propagate in opposite directions along the preferred axis.
This decomposition is not unique. We discuss flux-normalised and field-normalised decomposition in a systematic way, analyse the symmetry properties of the decomposition operators and 
use these symmetry properties to derive reciprocity theorems for the decomposed wave fields, for both types of normalisation. 
Based on the field-normalised reciprocity theorems, we derive representation theorems for decomposed wave fields. 
In particular we derive double- and single-sided Kirchhoff-Helmholtz integrals for forward and backward propagation of decomposed wave fields. The single-sided
Kirchhoff-Helmholtz integrals for backward propagation of field-normalised decomposed wave fields find applications in reflection imaging, accounting for multiple scattering.
\end{abstract}

\maketitle

\section{Introduction}\label{sec1}

In many wave propagation problems it is possible to define a preferred direction of propagation.
For example, in ocean acoustics, waves propagate primarily in the horizontal direction in an acoustic wave guide, bounded by the water surface and the ocean bottom. 
Similarly, in communication engineering,
microwaves or optical waves propagate as beams through electromagnetic or optical wave guides. 
Wave propagation in preferred directions is not restricted to wave guides. For example, in  geophysical \rev{reflection} imaging applications, 
seismic or electromagnetic waves propagate mainly in the vertical direction (downward and upward) through a laterally unbounded medium.

To account for propagation in preferred directions, 
the wave equation for the full wave field can be decomposed into a set of coupled equations for waves that propagate in opposite directions along the preferred axis
(for example leftward and rightward in ocean acoustics, or downward and upward in \rev{reflection} imaging). 
In the literature on electromagnetic wave propagation these oppositely propagating waves  are often called ``bidirectional beams''
%$^{1,2}$ 
\citep{Hoekstra97OQE, Stralen97OQE}
whereas in the acoustic literature they are usually called ``one-way wave fields''
%$^{3-7}$ 
\citep{Claerbout71GEO, Berkhout82Book, McCoy87WM,  Holberg88GP, Fishman93RS}. 
In this paper we use the latter terminology.

There is a vast amount of literature on the analytical and numerical aspects of one-way wave propagation
%$^{8-13}$ 
\citep{Fishman84JMP, Fishman84JMP2, Halpern88JASA, Weston89JMP, Fishman91WM, Grimbergen98GEO2}. 
A discussion of this is beyond the scope of this paper. Instead,
we concentrate on the choice of the decomposition operator and the consequences for reciprocity and representation theorems.
 
Decomposition of a wave field into one-way wave fields is not unique. In particular, the amplitudes of the one-way wave fields can be scaled in different ways. 
In this paper we distinguish between so-called ``flux-normalised'' and ``field-normalised'' one-way wave fields. 
The square of the amplitude of a flux-normalised one-way wave field is by definition the power-flux density 
(or, for quantum-mechanical waves, the probability-flux density)
in the direction of preference. 
Field-normalised one-way wave fields, on the other hand, are scaled such that the sum of the two oppositely propagating components equals the full wave field.
These two forms of normalisation have been briefly analysed by De Hoop
%$^{14,15}$ 
\citep{Hoop92PHD, Hoop96JMP}. 
From this analysis it appeared that the operators for flux-normalised decomposition
exhibit more symmetry than the operators for field-normalised decomposition. Exploiting the symmetry of the flux-normalised decomposition operators, the author derived reciprocity
and representation theorems for flux-normalised one-way wave fields
%$^{16,17}$ 
\citep{Wapenaar96GJI1, Wapenaar96GJI2}.

The first aim of this paper is to discuss flux-normalised versus field-normalised decomposition in a systematic way. 
In particular, it will be shown that reciprocity 
theorems  for field-normalised one-way wave fields can be derived in a similar way as those for flux-normalised one-way wave fields, 
even though the operators for field-normalised decomposition exhibit less symmetry.

The second aim is to discuss representation theorems for field-normalised one-way wave fields in a systematic way. This discussion includes links to ``classical'' Kirchhoff-Helmholtz 
integrals for one-way wave fields as well as to recent single-sided representations for backward propagation, used for example in Marchenko imaging
%$^{18}$ 
\citep{Wapenaar2014GEO}.
Despite the links to earlier results, the discussed representations are more general.  
An advantage of the representations for field-normalised one-way wave fields is that a straightforward summation of the one-way wave fields gives the full wave field.

We restrict the discussion to scalar wave fields. In section \ref{sec2} we formulate a unified scalar wave equation for acoustic waves,  
horizontally polarised shear waves, transverse electric and transverse magnetic EM waves
and, finally, quantum-mechanical waves.
Next, we reformulate the unified wave equation into a matrix-vector form, discuss symmetry properties of the operator matrix and use this to derive reciprocity theorems 
in matrix-vector form.
In section \ref{sec4} we decompose the matrix-vector wave equation into a coupled system of  equations for oppositely propagating one-way wave fields. We separately consider
flux-normalisation and field-normalisation  and derive reciprocity theorems for \rev{one-way wave fields, using} both normalisations.
In section \ref{sec5} we extensively discuss representation theorems for field-normalised one-way wave fields and indicate applications. We end with conclusions in section \ref{sec6}.

\section{Unified wave equation and its symmetry properties}\label{sec2}

\subsection{Unified scalar wave equation}

%%\rev{
%%%Assuming $u(t)$ is a real-valued function, equation (\ref{eqA11}) implies $u(-\omega)=u^*(\omega)$, where the asterisk denotes complex conjugation. Hence, since
%%The inverse Fourier transform is defined as
%%
%%\begin{eqnarray}\label{eqA11g}
%%&&u(t)=\frac{1}{2\pi}\int_{-\infty}^\infty u(\omega)\exp(-\i\omega t){\rm d}\omega.
%%\end{eqnarray}
%%
%%Assuming $u(t)$ is a real-valued function, its Fourier transform obeys the property $u(-\omega)=u^*(\omega)$, where the asterisk denotes complex conjugation.
%%Hence, the inverse Fourier transform can be replaced by
%%
%%\begin{eqnarray}\label{eqA11gg}
%%&&u(t)=\frac{1}{\pi}\Re\int_{0}^\infty u(\omega)\exp(-\i\omega t){\rm d}\omega,
%%\end{eqnarray}
%%
%%where $\Re$ denotes the real part. Since negative frequencies are not needed for the inverse Fourier transform, from here onward we con}

Using a unified notation, wave propagation in a lossless medium 
(or, for quantum-mechanical waves, in a lossless potential) 
is governed by the following two equations in the space-frequency  domain
\begin{eqnarray}
&&-\i\omega\alpha P+\partial_jQ_j=B,\label{eqA1}\\
&&-\i\omega\beta Q_j+\partial_j P=C_j.\label{eqA2}
\end{eqnarray}
\rev{Here $\i$ is the imaginary unit and $\omega$ the angular frequency (in this paper we consider positive frequencies only).
Operator $\partial_j$ stands for the spatial differential operator $\partial/\partial x_j$  and Einstein's summation convention applies to  repeated subscripts.}
 $P(\bx,\omega)$ and $Q_j(\bx,\omega)$ are space- and frequency-dependent wave field quantities, 
$\alpha(\bx)$ and $\beta(\bx)$ are  real-valued space-dependent parameters, and
$B(\bx,\omega)$ and $C_j(\bx,\omega)$ are space- and frequency-dependent source distributions.
\rev{Parameters $\alpha$ and $\beta$ are both assumed to be positive, hence, metamaterials are not considered in this paper.} 
All  quantities are specified in Table 1 for different wave phenomena and are discussed in more detail below.
As indicated in the first column of Table 1, we consider 3D and 2D wave problems. For the 3D situation, $\bx=(x_1,x_2,x_3)$ is the 3D Cartesian coordinate vector and 
\rev{lower-case} Latin subscripts take on the values 1, 2 and 3. For the 2D situation, $\bx=(x_1,x_3)$ is the 2D Cartesian coordinate vector and 
\rev{lower-case}  Latin subscripts take on the values 1 and 3 only.

\begin{center}
{{\noindent \it Table 1: Quantities in unified equations (\ref{eqA1}) and (\ref{eqA2}).}
\begin{tabular}
{||l||c|c||c|c||c|c||}
\hline\hline
& $P$ & $Q_j$  &$\alpha$ &$\beta$  & $B$ & $C_j$  \\
\hline\hline
1. Acoustic waves (3D)  & $p$ & $v_j$ &$\kappa$ &$\rho$  & $q$ & $f_j$ \\
\hline
2. SH waves (2D) & $v_2$ & $-\tau_{2j}$  &$\rho$ &$\frac{1}{\mu}$ & $f_2$ & $2h_{2j}$  \\
\hline
3. TE waves (2D) & $E_2$ & $-\epsilon_{2jk}H_k$  &$\varepsilon$ &$\mu$  &  $-J_2^{\rm e}$ & $\epsilon_{2jk}J_k^{\rm m}$  \\
\hline
4. TM waves (2D) & $H_2$ & $\epsilon_{2jk}E_k$ &$\mu$ &$\varepsilon$  & $-J_2^{\rm m}$ & $-\epsilon_{2jk}J_k^{\rm e}$  \\
\hline
5. Quantum waves (3D) & $\Psi$ & $\frac{2\hbar}{m\i}\partial_j\Psi$  &$4-\frac{4V}{\hbar\omega}$  &$\frac{m}{2\hbar\omega}$ &  &  \\
\hline
\hline
\end{tabular}
}
\end{center}

\mbox{}\\

\rev{The unified boundary conditions at an interface  between two media with different parameters state that $P$ and $n_jQ_j$ are continuous over the interface. Here $n_j$ represents the
 components of the normal vector ${\bf n}=(n_1,n_2,n_3)$ at the interface for the 3D situation, or ${\bf n}=(n_1,n_3)$ for the 2D situation.}
 
We discuss the quantities in Table 1 in more detail. The quantities in row 1, associated to 3D acoustic wave propagation in a lossless fluid medium, are 
acoustic pressure $p(\bx,\omega)$, particle velocity $v_j(\bx,\omega)$,  compressibility $\kappa(\bx)$, mass density $\rho(\bx)$, 
volume-injection rate density $q(\bx,\omega)$ and external force density $f_j(\bx,\omega)$.
For 2D horizontally polarised shear waves in a lossless solid medium, we have in row 2 horizontal particle velocity $v_2(\bx,\omega)$, shear stress $\tau_{2j}(\bx,\omega)$,
mass density $\rho(\bx)$, shear modulus $\mu(\bx)$, external horizontal force density $f_2(\bx,\omega)$ and external shear deformation rate density $h_{2j}(\bx,\omega)$.
Rows 3 and 4 contain the quantities for 2D electromagnetic wave propagation, with TE standing for transverse electric and TM for transverse magnetic. The quantities are
electric field strength $E_k(\bx,\omega)$, magnetic field strength $H_k(\bx,\omega)$, permittivity $\varepsilon(\bx)$, permeability $\mu(\bx)$,
external electric current density $J_k^{\rm e}(\bx,\omega)$ and external magnetic current density $J_k^{\rm m}(\bx,\omega)$. Furthermore, $\epsilon_{ijk}$ is
the alternating tensor (or Levi-Civita tensor), with $\epsilon_{123}=\epsilon_{312}=\epsilon_{231}=1$,
$\epsilon_{213}=\epsilon_{321}=\epsilon_{132}=-1$, and all other components being zero. 
In row 5, the quantities related to 3D quantum-mechanical wave propagation are wave function $\Psi (\bx,\omega)$, 
potential $V(\bx)$, particle mass $m$ and $\hbar=h/2\pi$, with $h$ Planck's constant.

By eliminating $Q_j$ from equations (\ref{eqA1}) and (\ref{eqA2}) we obtain the unified  scalar wave equation
\begin{eqnarray}
&&\beta\partial_j\Bigl(\frac{1}{\beta}\partial_j P\Bigr) +k^2P=\beta\partial_j\Bigl(\frac{1}{\beta}C_j\Bigr)+\i\omega\beta B,\label{eqwe}
\end{eqnarray}
with wave number $k$ defined via
\begin{eqnarray}\label{eqk}
&&k^2=\alpha\beta\omega^2.
\end{eqnarray}

\subsection{Unified wave equation in matrix-vector form}

\begin{figure}
\vspace{0cm}
\centerline{\epsfysize=8 cm \epsfbox{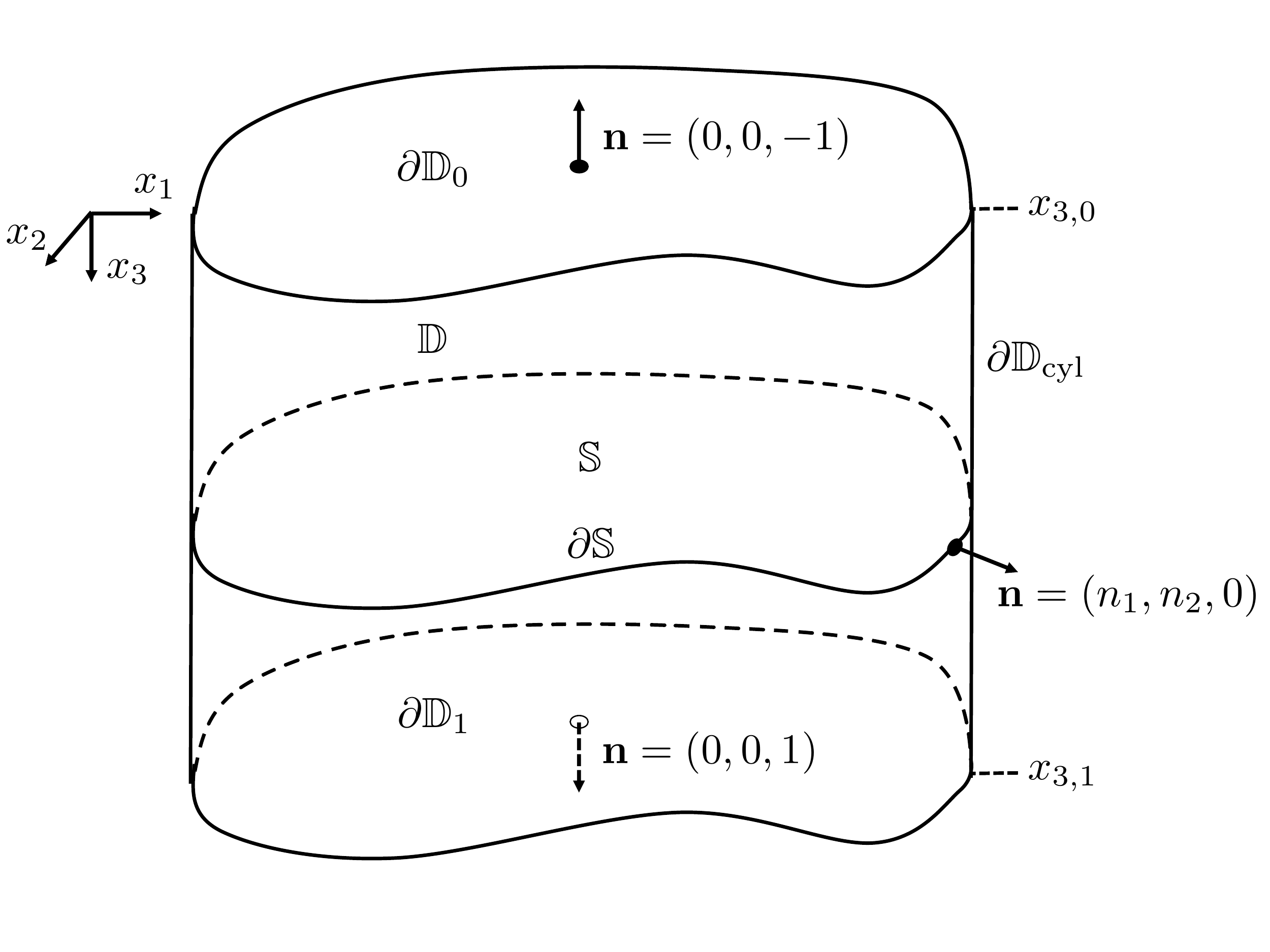}}
%\vspace{-2.8cm}
\caption{\footnotesize 
\rev{Configuration with the $x_3$-direction as the preferred direction. In the lateral direction this configuration can be bounded (for wave guides) or unbounded (for example for geophysical reflection imaging applications).
For the 2D situation the configuration is a cross-section of the 3D situation for $x_2=0$.}
}\label{Fig1}
\end{figure}

\rev{We define a configuration with a preferred direction and reorganise equations (\ref{eqA1}) and (\ref{eqA2}) accordingly.}

\rev{Consider a 3D spatial domain $\setD$, enclosed by surface $\setdD$. 
This surface consists of two planar surfaces $\setdD_0$ and $\setdD_1$ perpendicular to the $x_3$-axis and a cylindrical surface $\setdD_{\rm cyl}$
with its axis parallel to the $x_3$-axis, see Figure \ref{Fig1}. 
The surfaces $\setdD_0$ and $\setdD_1$ are situated 
 at $x_3=x_{3,0}$ and $x_3=x_{3,1}$,  respectively, with $x_{3,1}> x_{3,0}$.  In general these surfaces do not coincide with  physical boundaries.
 Surface $\setA$ in Figure \ref{Fig1} is a cross-section of $\setD$ at arbitrary $x_3$.
 The parameters $\alpha(\bx)$ and $\beta(\bx)$ are piecewise continuous smoothly varying functions in $\setD$, with discontinuous jumps only at interfaces that are perpendicular to the $x_3$-axis
 (hence, $P$ and $Q_3$ are continuous over the interfaces).
 In the lateral direction the domain $\setD$ can be bounded or unbounded. When $\setD$ is laterally bounded, the configuration in Figure \ref{Fig1} represents a wave guide. 
% When $\setD$ is a wave guide, the cylindrical surface $\setdD_{\rm cyl}$ is finite.
 For this situation we assume that homogeneous Dirichlet or Neumann boundary conditions apply, i.e., $P=Q_3=0$ or $n_\nu\partial_\nu P=n_\nu\partial_\nu Q_3=0$ at $\setdD_{\rm cyl}$,
where lower-case Greek subscripts take on the values 1 and 2.
 When $\setD$ is laterally unbounded (for example for reflection imaging applications), the cylindrical surface $\setdD_{\rm cyl}$ has an infinite radius and we assume that 
 $P$ and $Q_3$ have ``sufficient decay''  at infinity.
 For the 2D situation, the configuration is a cross-section of the 3D situation for $x_2=0$ and lower-case Greek subscripts take on the value 1 only.}
% The two surfaces $\setdD_0$ and $\setdD_1$ are together denoted by $\setdD$. In general $\setdD$ does not coincide with a physical boundary.}

We reorganise  equations (\ref{eqA1}) and (\ref{eqA2}) into a matrix-vector wave equation which acknowledges  \rev{the $x_3$-direction as the} direction of preference.
By eliminating the lateral components $Q_1$ and $Q_2$ (or, for 2D wave problems, the lateral component $Q_1$),
we obtain
%$^{8,15,19-21}$ 
\citep{Corones75JMAA, Ursin83GEO, Fishman84JMP, Wapenaar89Book, Hoop96JMP}
\begin{eqnarray}\label{eq2.1}
&&\partial_3\bq=\bA\bq+\bd,
\end{eqnarray}
where wave vector $\bq$ and source vector $\bd$ are defined as
\begin{eqnarray}
&&\bq=
\begin{pmatrix} P\\ Q_3\end{pmatrix},
\quad \bd=
\begin{pmatrix}C_3\\ B_0\end{pmatrix}, \label{eq2.2q}
\end{eqnarray}
with
\begin{eqnarray}
&&B_0=B+\frac{1}{\i\omega}\partial_\nu\frac{1}{\beta}C_\nu\label{eq2.5}
\end{eqnarray}
and operator matrix $\bA$ defined as
\begin{eqnarray}
&&\quad\bA=\begin{pmatrix} 0 & {\cal A}_{12} \\ {\cal A}_{21} & 0\end{pmatrix}, \label{eq2.2}
\end{eqnarray}
with
\begin{eqnarray}
&&{\cal A}_{12}=\i\omega\beta,\label{eq2.3}\\
&&{\cal A}_{21}=\i\omega\alpha-\frac{1}{\i\omega}\partial_\nu\frac{1}{\beta}\partial_\nu.\label{eq2.4}
\end{eqnarray}
The notation in the right-hand side of equations (\ref{eq2.5}) and (\ref{eq2.4}) should be understood in the sense that differential operators 
act on all factors to the right of it.  Hence, operator $\partial_\nu\frac{1}{\beta}\partial_\nu$, 
applied via equation (\ref{eq2.1}) to $P$, stands for $\partial_\nu (\frac{1}{\beta}\partial_\nu P)$. 

Note that the quantities contained in the wave vector $\bq$ \rev{are continuous over interfaces perpendicular to the $x_3$-axis. Moreover, these quantities }constitute the power-flux density 
(or, for quantum-mechanical waves, the probability-flux density)
in the $x_3$-direction via
\begin{eqnarray}\label{eqJ}
&&\J=\quart\{P^*Q_3+Q_3^*P\}.
\end{eqnarray}
where the asterisk denotes complex conjugation.

\subsection{Symmetry properties of the operator matrix}\label{secsymx}

We discuss the symmetry properties of the operator matrix $\bA$.
First, consider a general operator ${\cal U}$ (which can be a scalar or a matrix), containing space-dependent parameters  and differential operators $\partial_\nu$.
We introduce the transpose operator ${\cal U}^t$ via the following integral property
\begin{eqnarray}\label{eq90c}
&&\int_\setA ({\cal U}f)^tg\,\rmd\bxh=\int_\setA f^t({\cal U}^tg)\,\rmd\bxh.
\end{eqnarray}
Here $\bxh$ is the lateral coordinate vector, with $\bxh=(x_1,x_2)$  for 3D and $\bxh=x_1$ for 2D wave problems.
$\setA$ denotes an  integration surface perpendicular to the $x_3$-axis at arbitrary $x_3$, \rev{with edge $\partial\setA$, see Figure \ref{Fig1}.}
The quantities $f(\bxh)$ and $g(\bxh)$ are space-dependent test functions (scalars or vectors).
\rev{When these functions are vectors, $f^t$ is the transpose of $f$; when they are scalars, $f^t$ is equal to $f$.}
\rev{When $\setA$ is bounded, homogeneous Dirichlet or Neumann conditions are imposed at  $\partial\setA$.
When $\setA$ is unbounded,  $\partial\setA$ has an infinite radius and $f(\bxh)$ and $g(\bxh)$ are assumed to have} sufficient decay along $\setA$ towards infinity. 
Operator ${\cal U}$ is said to be symmetric when ${\cal U}^t={\cal U}$ and skew-symmetric when ${\cal U}^t=-{\cal U}$.
For the special case that ${\cal U}=\partial_\nu$, equation  (\ref{eq90c}) implies $\partial_\nu^t=-\partial_\nu$ (via integration by parts \rev{along $\setA$}). Hence, operator $\partial_\nu$ is skew-symmetric.

We introduce the adjoint operator ${\cal U}^\dagger$ (\rev{i.e., the complex conjugate transpose of ${\cal U}$}) 
via the integral property
\begin{eqnarray}\label{eq90d}
&&\int_\setA ({\cal U}f)^\dagger g\,\rmd\bxh=\int_\setA f^\dagger({\cal U}^\dagger g)\,\rmd\bxh.
\end{eqnarray}
\rev{When the test functions are vectors, $f^\dagger$ is the complex conjugate transpose of $f$; when they are scalars, $f^\dagger$ is the complex conjugate of $f$.}
Operator ${\cal U}$ is said to be Hermitian (or self-adjoint) when ${\cal U}^\dagger={\cal U}$ and skew-Hermitian when ${\cal U}^\dagger=-{\cal U}$.
For the operators ${\cal A}_{12}$ and ${\cal A}_{21}$, defined in equations (\ref{eq2.3}) and (\ref{eq2.4}), we find ${\cal A}_{12}^t={\cal A}_{12}$, ${\cal A}_{21}^t={\cal A}_{21}$,
${\cal A}_{12}^\dagger=-{\cal A}_{12}$ and ${\cal A}_{21}^\dagger=-{\cal A}_{21}$. 
Hence, operators ${\cal A}_{12}$ and ${\cal A}_{21}$ are symmetric and skew-Hermitian.
With these relations, we find for the operator matrix $\bA$
\begin{eqnarray}
\bA^t\bN&=&-\bN\bA,\label{eqsym}\\
\bA^\dagger\bK&=&-\bK\bA,\label{eqsymad}
\end{eqnarray}
with
\begin{eqnarray}\label{eq4.3NK}
&&{\bN}=\begin{pmatrix} 0 & 1 \\ -1 & 0 \end{pmatrix},
\quad {\bK}=\begin{pmatrix} 0 & 1 \\ 1 & 0 \end{pmatrix}.
\end{eqnarray}
Note that, using the expressions for $\bq$ and $\bK$ in equations (\ref{eq2.2q}) and (\ref{eq4.3NK}), we can rewrite equation (\ref{eqJ}) for the 
power-flux density 
(or, for quantum-mechanical waves, the probability-flux density)
as
\begin{eqnarray}
&&\J=\quart\bq^\dagger\bK\bq.\label{eqjj}
\end{eqnarray}

\subsection{Reciprocity theorems}\label{sec3d}

We derive reciprocity theorems between two independent solutions of wave equation (\ref{eq2.1}) \rev{for the configuration of Figure \ref{Fig1}.}
We consider two states $A$ and $B$, characterised by wave vectors $\bq_A(\bx,\omega)$ and $\bq_B(\bx,\omega)$, obeying wave equation (\ref{eq2.1}),
with source vectors $\bd_A(\bx,\omega)$ and $\bd_B(\bx,\omega)$. 
In domain $\setD$, the parameters $\alpha$ and $\beta$, and hence the matrix operator $\bA$, are chosen the same in the two states 
(outside $\setdD$ they may be different in the two states).
Consider the quantity $\partial_3\bigl(\bq_A^t\bN\bq_B\bigr)$ in domain $\setD$. 
Applying the product rule for differentiation, using equation (\ref{eq2.1}) for both states,
integrating the result over $\setD$ and applying the theorem of Gauss yields
\begin{eqnarray}\label{eq4.10}
&&\int_\setD\Bigl(\bigl((\bA\bq_A)^t+\bd_A^t\bigr)\bN\bq_B +\bq_A^t\bN\bigl(\bA\bq_B+\bd_B\bigr) \Bigr)\rmdd\bx=\int_\setdD\bq_A^t\bN\bq_B n_3\rmd\bx.
\end{eqnarray}
Here $n_3$ is the component parallel to the $x_3$-axis of the outward pointing normal vector on $\setdD$, with $n_3=-1$ at $\setdD_0$, 
$n_3=+1$ at $\setdD_1$ \rev{and $n_3=0$ at $\setdD_{\rm cyl}$, see Figure \ref{Fig1}. In the following the integral on the right-hand side is restricted to the horizontal surfaces $\setdD_0$ and $\setdD_1$, which
together are denoted by $\setdDD$.} The integral on the left-hand side can be written as $\int_\setD(\cdots)\rmdd\bx=\int_{x_{3,0}}^{x_{3,1}}{\rm d}x_3\int_\setA(\cdots)\rmd\bxh$.
Using equation (\ref{eq90c}) for the integral along $\setA$ and symmetry property (\ref{eqsym}), it follows that the two terms in equation (\ref{eq4.10}) containing operator $\bA$ cancel each other. 
Hence, we are left with
\begin{eqnarray}\label{eq4.1}
&&\int_\setD\bigl(\bd_A^t\bN\bq_B +\bq_A^t\bN\bd_B \bigr)\rmdd\bx=
\int_\setdDD\bq_A^t\bN\bq_B n_3\rmd\bxh.
\end{eqnarray}
This is a convolution-type reciprocity theorem
%$^{22-24}$ 
\citep{Fokkema93Book, Hoop95Book, Achenbach2003Book}, 
because products like $\bq_A^t(\bx,\omega)\bN\bq_B(\bx,\omega)$ in the frequency domain correspond to convolutions in the time domain. 
A more familiar form is obtained by substituting the expressions for $\bq$, $\bd$ and $\bN$ (equations \ref{eq2.2q} and \ref{eq4.3NK}), choosing $C_j=0$ and using equation
 (\ref{eqA2}) to eliminate $Q_3$, which gives
\begin{eqnarray}\label{eq4.1comp}
&&\int_\setD(-B_AP_B+P_A B_B)\rmdd\bx=
\int_\setdDD\frac{1}{\i\omega\beta}\bigl(P_A \partial_3P_B-(\partial_3P_A)P_B\bigr)n_3\rmd\bxh.
\end{eqnarray}
Next, consider the quantity $\partial_3\bigl(\bq_A^\dagger\bK\bq_B\bigr)$ in domain $\setD$. Following the same steps as above, using equations (\ref{eq90d}) and (\ref{eqsymad})
instead of (\ref{eq90c}) and (\ref{eqsym}), we obtain
\begin{eqnarray}\label{eq4.2}
&&\int_\setD\bigl(\bd_A^\dagger\bK\bq_B + \bq_A^\dagger\bK\bd_B\bigr)\rmdd\bx=
\int_\setdDD\bq_A^\dagger\bK\bq_B n_3\rmd\bxh.
\end{eqnarray}
This is a correlation-type reciprocity theorem
%$^{25}$ 
\citep{Bojarski83JASA}, 
because products like $\bq_A^\dagger(\bx,\omega)\bK\bq_B(\bx,\omega)$ in the frequency domain correspond to correlations in the time domain.
Substituting the expressions for $\bq$, $\bd$ and $\bK$ and choosing $C_j=0$ yields the more familiar form
\begin{eqnarray}\label{eq4.2comp}
&&\int_\setD(B_A^*P_B+P_A^* B_B)\rmdd\bx=
\int_\setdDD\frac{1}{\i\omega\beta}\bigl(P_A^* \partial_3P_B-(\partial_3P_A)^*P_B\bigr)n_3\rmd\bxh.
\end{eqnarray}
We obtain a  special case by choosing states $A$ and $B$ identical. Dropping the subscripts $A$ and $B$ in equations (\ref{eq4.2}) and (\ref{eq4.2comp}) 
\rev{and multiplying both sides of these equations by $\quart$,} gives
\begin{eqnarray}\label{eq4.3}
&&\frac{1}{4}\int_\setD\bigl(\bd^\dagger\bK\bq + \bq^\dagger\bK\bd\bigr)\rmdd\bx=
\frac{1}{4}\int_\setdDD\bq^\dagger\bK\bq n_3\rmd\bxh
\end{eqnarray}
and
\begin{eqnarray}\label{eq4.3comp}
&&\frac{1}{4}\int_\setD(B^*P+P^* B)\rmdd\bx=
\frac{1}{4}\int_\setdDD\frac{1}{\i\omega\beta}\bigl(P^* \partial_3P-(\partial_3P)^*P\bigr)n_3\rmd\bxh,
\end{eqnarray}
respectively. These equations quantify conservation of power
(or, for quantum-mechanical waves, probability).

\section{Decomposed wave equation and its symmetry properties}\label{sec4}

\subsection{General decomposition of the matrix-vector wave equation}

To facilitate the decomposition of the matrix-vector wave equation (equation \ref{eq2.1}), 
we recast the operator matrix $\bA$ into a somewhat different form.
To this end we introduce an operator ${\cal H}_2$, according to
\begin{eqnarray}
{\cal H}_2&=&-\i\omega\sqrt{\beta}{\cal A}_{21}\sqrt{\beta}\nonumber\\
&=& k^2+\sqrt{\beta}\partial_\nu\frac{1}{\beta}\partial_\nu\sqrt{\beta},\label{eq39}
\end{eqnarray}
with operator ${\cal A}_{21}$ defined in equation (\ref{eq2.4}) and wavenumber  $k$ in equation (\ref{eqk}).
Operator ${\cal H}_2$ can be rewritten as a Helmholtz operator
%$^{14,21}$ 
\citep{Wapenaar89Book, Hoop92PHD}
\begin{eqnarray}\label{eqAp7}
&&{\cal H}_2= k_s^2 + \partial_\nu\partial_\nu,
% \quad k_s^2=\frac{\omega^2}{c^2}-\frac{3(\partial_\nu\beta)(\partial_\nu\beta)}{4\beta^2}+\frac{\partial_\nu\partial_\nu\beta}{2\beta}.
\end{eqnarray}
 with the scaled wavenumber $k_s$ defined as
 %$^{26}$ 
\citep{Brekhovskikh60Book}
\begin{eqnarray}
&&k_s^2=k^2-\frac{3(\partial_\nu\beta)(\partial_\nu\beta)}{4\beta^2}+\frac{(\partial_\nu\partial_\nu\beta)}{2\beta}.
\end{eqnarray}
Note that ${\cal H}_2^t={\cal H}_2$ and ${\cal H}_2^\dagger={\cal H}_2$, hence operator ${\cal H}_2$ is symmetric and self-adjoint \rev{and its spectrum is real-valued (with positive and negative eigenvalues)}.
Using equation (\ref{eq39}), we rewrite operator matrix $\bA$, defined in equation (\ref{eq2.2}), as
\begin{eqnarray}
&&\bA=\begin{pmatrix} 0 & \i\omega\beta\\ -\frac{1}{\i\omega\sqrt{\beta}}{\cal H}_2\frac{1}{\sqrt{\beta}} & 0\end{pmatrix}.\label{eqAmod}
\end{eqnarray}
Next, we decompose this operator matrix as follows
\begin{eqnarray}\label{eqdecom1}
&&\bA=\bL\bH\bL^{-1},
\end{eqnarray}
with
\begin{eqnarray}
\bH&=&\begin{pmatrix} \i\HH_1 & 0 \\  0 & -\i\HH_1\end{pmatrix},\label{eqH1}\\
\bL&=&\begin{pmatrix} {\cal L}_1 &  {\cal L}_1 \\  {\cal L}_2 & - {\cal L}_2\end{pmatrix},\label{eqL}\\
\bL^{-1}&=&\frac{1}{2}\begin{pmatrix} {\cal L}_1^{-1} &  {\cal L}_2^{-1} \\  {\cal L}_1^{-1} & - {\cal L}_2^{-1}\end{pmatrix}.\label{eqLi}
\end{eqnarray}
Operators $\HH_1$, ${\cal L}_1$ and ${\cal L}_2$ are pseudo-differential operators
%$^{7,8,14,16,21,27-30}$ 
\citep{Fishman84JMP, Fishman93RS, Wapenaar89Book, Hoop92PHD, Hoop94WM, Wapenaar96GJI1, Fishman2000JMP, Thomson2015GJI1, Thomson2015GJI2}.
The decomposition expressed by equation (\ref{eqdecom1}) is not unique, hence, different choices for operators $\HH_1$, ${\cal L}_1$ and ${\cal L}_2$ are possible. 
We discuss two of these choices in detail in the next two sections.
Here we derive some general relations that are independent of these choices.

By substituting equations (\ref{eqAmod}) and  (\ref{eqH1}) $-$ (\ref{eqLi}) into equation (\ref{eqdecom1}) we obtain the following relations
\begin{eqnarray}
\omega\beta &=& {\cal L}_1\HH_1{\cal L}_2^{-1},\label{eq34}\\
\frac{1}{\omega\sqrt{\beta}}{\cal H}_2\frac{1}{\sqrt{\beta}} &=& {\cal L}_2\HH_1{\cal L}_1^{-1}.\label{eq35}
\end{eqnarray}
We introduce a decomposed field vector $\bp$ and a decomposed source vector $\bs$ via
\begin{eqnarray}
&&\bq=\bL\bp, \quad \bp=\bL^{-1}\bq,\label{eqdecom2}\\
&&\bd=\bL\bs, \quad \bs=\bL^{-1}\bd,\label{eqdecom3}
\end{eqnarray}
where
\begin{eqnarray}
&& \bp=\begin{pmatrix} \p^+ \\ \p^- \end{pmatrix},\quad \bs=\begin{pmatrix} \s^+ \\ \s^- \end{pmatrix}.\label{eqdecom4}
\end{eqnarray}
Substitution of equations  (\ref{eqdecom1}),  (\ref{eqdecom2}) and  (\ref{eqdecom3}) into the matrix-vector wave equation (\ref{eq2.1}) yields
\begin{eqnarray}
&&\partial_3\bp=\bigl(\bH-\bL^{-1}\partial_3\bL\bigr)\bp+\bs.\label{eqoneway}
\end{eqnarray}
Substituting equations  (\ref{eqH1}) $-$ (\ref{eqLi}) and (\ref{eqdecom4}) into equation (\ref{eqoneway}) gives
\begin{eqnarray}
&&\hspace{-1.1cm}\partial_3\begin{pmatrix} \p^+ \\ \p^- \end{pmatrix}=\begin{pmatrix}  \i{\cal H}_1 & 0\\ 0 & -\i{\cal H}_1\end{pmatrix}\begin{pmatrix} \p^+ \\ \p^- \end{pmatrix}
-\frac{1}{2}\begin{pmatrix} {\cal L}_1^{-1} &  {\cal L}_2^{-1} \\  {\cal L}_1^{-1} & - {\cal L}_2^{-1}\end{pmatrix}
\begin{pmatrix} \partial_3{\cal L}_1 &  \partial_3{\cal L}_1 \\  \partial_3{\cal L}_2 & - \partial_3{\cal L}_2\end{pmatrix}
\begin{pmatrix} \p^+ \\ \p^- \end{pmatrix}+\begin{pmatrix} \s^+ \\ \s^- \end{pmatrix}.\label{eqAAA26}
\end{eqnarray}
This is a system of coupled one-way wave equations. From the first term on the right-hand side it follows that the one-way wave fields   
$\p^+$ and $\p^-$ propagate in the positive and negative $x_3$-direction, respectively. The second term on the right-hand side accounts for coupling between $\p^+$ and $\p^-$.
The last term on the right-hand side contains sources $\s^+$ and $\s^-$ which emit waves in the  positive and negative $x_3$-direction, respectively.

We conclude this section by substituting equations (\ref{eqdecom2}) and (\ref{eqdecom3}) into equations (\ref{eq4.1}), (\ref{eq4.2}) and (\ref{eq4.3}). 
Using equations (\ref{eq90c}) and (\ref{eq90d}) for the integration along
the lateral coordinates this yields
\begin{eqnarray}
\int_\setD\bigl(\bs_A^t\bL^t\bN\bL\bp_B +\bp_A^t\bL^t\bN\bL\bs_B \bigr)\rmdd\bx&=&\int_\setdDD\bp_A^t\bL^t\bN\bL\bp_B n_3\rmd\bxh,\label{eq41}\\
\int_\setD\bigl(\bs_A^\dagger\bL^\dagger\bK\bL\bp_B +\bp_A^\dagger\bL^\dagger\bK\bL\bs_B \bigr)\rmdd\bx&=&\int_\setdDD\bp_A^\dagger\bL^\dagger\bK\bL\bp_B n_3\rmd\bxh,\label{eq42}\\
\frac{1}{4}\int_\setD\bigl(\bs^\dagger\bL^\dagger\bK\bL\bp +\bp^\dagger\bL^\dagger\bK\bL\bs \bigr)\rmdd\bx&=&\frac{1}{4}\int_\setdDD\bp^\dagger\bL^\dagger\bK\bL\bp n_3\rmd\bxh.\label{eq43}
\end{eqnarray}
These equations form the basis for reciprocity theorems for the decomposed field and source vectors $\bp$ and $\bs$ in the next two sections.
 
\subsection{Flux-normalised decomposition and reciprocity theorems}

The first choice of operators $\HH_1$, ${\cal L}_1$ and ${\cal L}_2$ obeying equations (\ref{eq34}) and (\ref{eq35}) is
%$^{14-16}$ 
\citep{Hoop92PHD, Hoop96JMP, Wapenaar96GJI1}
\begin{eqnarray}
\HH_1&=&\HH_2^{\half},\\
{\cal L}_1&=&(\omega/2)^\half\beta^{\half}\HH_1^{-\half},\label{eqB19q}\\
{\cal L}_2&=&(2\omega)^{-\half}\beta^{-\half}\HH_1^{\half}.\label{eqB20q}
\end{eqnarray}
Operator $\HH_1$, which is the square root of the Helmholtz operator $\HH_2$, is commonly known as the square-root operator
%$^{3,4,8}$  
\citep{Claerbout71GEO, Berkhout82Book, Fishman84JMP}.
Like the Helmholtz operator $\HH_2$, the square-root operator $\HH_1$  is a symmetric operator
%$^{16}$ 
\citep{Wapenaar96GJI1}, 
hence $\HH_1^t=\HH_1$. 
For the adjoint square-root operator we have $\HH_1^\dagger=(\HH_1^t)^*=\HH_1^*$. The spectrum of $\HH_1$ is real-valued for propagating waves and imaginary-valued for evanescent waves.
Hence, unlike the Helmholtz operator, the square-root operator is not self-adjoint. If we neglect evanescent waves, we may approximate the adjoint 
square-root operator as $\HH_1^\dagger\approx\HH_1$. Similar relations hold for the square root of the square-root operator and its inverse, hence 
$\bigl(\HH_1^{\pm \half}\bigr)^t=\HH_1^{\pm \half}$ and, neglecting evanescent waves, $\bigl(\HH_1^{\pm \half}\bigr)^\dagger\approx\HH_1^{\pm \half}$. 
%\rev{Although neglecting evanescent waves implies that several symmetry relations are not exact, an advantage is that expressions for backward propagation,
%such as those derived in sections \ref{sec4d} and \ref{sec4e}, are stable.}
From here onward we replace $\approx$ by $=$ when the only approximation is the negligence of evanescent waves.
Using these symmetry relations for $\HH_1$ and equations (\ref{eq4.3NK}), (\ref{eqL}), (\ref{eqB19q}) and (\ref{eqB20q}), we obtain
\begin{eqnarray}
&&\bL^t\bN\bL=-\bN\label{eqA32f}
\end{eqnarray}
and, neglecting evanescent waves,
\begin{eqnarray}
&&\bL^\dagger\bK\bL=\bJ,\label{eqA33f}
\end{eqnarray}
with
\begin{eqnarray}\label{eq4.3J}
&&{\bJ}=\begin{pmatrix} 1 & 0 \\ 0 & -1 \end{pmatrix}.
\end{eqnarray}
Hence, equations (\ref{eq41}) $-$ (\ref{eq43}) simplify to
\begin{eqnarray}
-\int_\setD\bigl(\bs_A^t\bN\bp_B +\bp_A^t\bN\bs_B \bigr)\rmdd\bx&=&-\int_\setdDD\bp_A^t\bN\bp_B n_3\rmd\bxh,\\
\int_\setD\bigl(\bs_A^\dagger\bJ\bp_B +\bp_A^\dagger\bJ\bs_B \bigr)\rmdd\bx&=&\int_\setdDD\bp_A^\dagger\bJ\bp_B n_3\rmd\bxh,\\
\frac{1}{4}\int_\setD\bigl(\bs^\dagger\bJ\bp +\bp^\dagger\bJ\bs \bigr)\rmdd\bx&=&\frac{1}{4}\int_\setdDD\bp^\dagger\bJ\bp n_3\rmd\bxh.
\end{eqnarray}
By substituting the expressions for $\bp$, $\bs$, $\bN$ and $\bJ$ (equations \ref{eqdecom4}, \ref{eq4.3NK} and \ref{eq4.3J}), we obtain
\begin{eqnarray}
\int_\setD(-\s_A^+\p_B^- + \s_A^-\p_B^+-\p_A^+\s_B^- +\p_A^-\s_B^+)\rmdd\bx&=&
\int_\setdDD(-\p_A^+\p_B^- +\p_A^-\p_B^+)n_3\rmd\bxh,\label{eq4.1deccomp}\\
\int_\setD(\s_A^{+*}\p_B^+ -\s_A^{-*}\p_B^- + \p_A^{+*}\s_B^+ -\p_A^{-*}\s_B^-)\rmdd\bx&=&
\int_\setdDD(\p_A^{+*}\p_B^+ -\p_A^{-*}\p_B^-)n_3\rmd\bxh,\label{eq4.2deccomp}\\
\frac{1}{4}\int_\setD(\s^{+*}\p^+ -\s^{-*}\p^- + \p^{+*}\s^+ -\p^{-*}\s^-)\rmdd\bx
&=&\frac{1}{4}\int_\setdDD(|\p^+|^2 -|\p^-|^2)n_3\rmd\bxh.\label{eq4.3deccomp}
\end{eqnarray}
Note that, since the right-hand side of equation (\ref{eq4.3deccomp}) is equal to the right-hand side of equation (\ref{eq4.3comp}), it quantifies the power flux  
(or the probability-flux for quantum-mechanical waves)
through the surface $\setdDD$. Therefore we call $\p^+$ and $\p^-$ flux-normalised one-way wave fields. 
Consequently, equations (\ref{eq4.1deccomp}) and (\ref{eq4.2deccomp}) are reciprocity theorems of the convolution type and correlation type, 
respectively, for flux-normalised one-way wave fields.
These theorems have been derived previously
%$^{16}$ 
\citep{Wapenaar96GJI1} 
and have found applications in advanced wave field imaging methods for active and passive data
%$^{31-42}$
\citep{Wapenaar2004GJI, Kumar2006JGR, Fan2006GEO, Weglein2006GEO, Slob2009IEEE, Ruigrok2012GRL, Wapenaar2014JASA, Slob2014GEO, Ravasi2016GJI, Ravasi2017GEO, Elison2018GJI, Staring2018GEO}.

\subsection{Field-normalised decomposition and reciprocity theorems}\label{sec3c}

The second choice of operators $\HH_1$, ${\cal L}_1$ and ${\cal L}_2$ obeying equations (\ref{eq34}) and (\ref{eq35}) is
%$^{21}$ 
\citep{Wapenaar89Book}
\begin{eqnarray}
\HH_1&=&\beta^\half\HH_2^{\half}\beta^{-\half},\label{eqB18qf}\\
{\cal L}_1&=&1,\label{eqB19qf}\\
{\cal L}_2&=&(\omega\beta)^{-1}\HH_1.\label{eqB20qf}
\end{eqnarray}
Only the Helmholtz operator $\HH_2$ is the same as in the previous section (it is defined in equation \ref{eqAp7}). 
The operators $\HH_1$, ${\cal L}_1$ and ${\cal L}_2$ are different from those in the previous section, but for convenience we use the same symbols.
Using $\bq=\bL\bp$ (equation \ref{eqdecom2}) and equations (\ref{eq2.2q}), (\ref{eqL}), (\ref{eqdecom4}) and (\ref{eqB19qf}), we find
\begin{eqnarray}
&&P=\p^+ + \p^-,\label{eq59}
\end{eqnarray}
hence, $\p^+$ and $\p^-$ have the same physical dimension as the full field variable $P$ (which is defined in Table 1 for different wave phenomena).
Therefore we call $\p^+$ and $\p^-$ field-normalised one-way wave fields (for convenience we use the same symbols as in the previous section). 

The square-root operator $\HH_2^\half$ is symmetric, but $\HH_1$ defined in equation (\ref{eqB18qf}) is not. 
From this equation it easily follows that $\HH_1$, premultiplied by $\beta^{-1}$ is symmetric, hence
\begin{eqnarray}
\biggl(\frac{1}{\beta}\HH_1\biggr)^t &=&\frac{1}{\beta} \HH_1
\end{eqnarray}
and, neglecting evanescent waves,
\begin{eqnarray}
\biggl(\frac{1}{\beta}\HH_1\biggr)^\dagger &=&
\frac{1}{\beta} \HH_1.
\end{eqnarray}
Using these symmetry relations for $\frac{1}{\beta}\HH_1$ and equations (\ref{eq4.3NK}), (\ref{eqL}), (\ref{eqB19qf}) and (\ref{eqB20qf}), we obtain
\begin{eqnarray}
&&\bL^t\bN\bL=
\begin{pmatrix} 0 & -2{\cal L}_2 \\ 2{\cal L}_2 & 0 \end{pmatrix}=-\bN\biggl(\frac{2}{\omega\beta}\HH_1\biggr)=-\biggl(\frac{2}{\omega\beta}\HH_1\biggr)^t\bN
\end{eqnarray}
and, neglecting evanescent waves,
\begin{eqnarray}
&&\bL^\dagger\bK\bL=
\begin{pmatrix}  2{\cal L}_2 & 0 \\ 0 & -2{\cal L}_2  \end{pmatrix}=\bJ\biggl(\frac{2}{\omega\beta}\HH_1\biggr)=\biggl(\frac{2}{\omega\beta}\HH_1\biggr)^\dagger\bJ.
\end{eqnarray}
Using this in equations (\ref{eq41}) and (\ref{eq42}) yields
\begin{eqnarray}
-\int_\setD\biggl[\bs_A^t\biggl(\frac{2}{\omega\beta}\HH_1\biggr)^t\bN\bp_B +\bp_A^t\bN\biggl(\frac{2}{\omega\beta}\HH_1\biggr)\bs_B \biggr]\rmdd\bx
&=&-\int_\setdDD\bp_A^t\biggl(\frac{2}{\omega\beta}\HH_1\biggr)^t\bN\bp_B n_3\rmd\bxh,\\
\int_\setD\biggl[\bs_A^\dagger\biggl(\frac{2}{\omega\beta}\HH_1\biggr)^\dagger\bJ\bp_B +\bp_A^\dagger\bJ\biggl(\frac{2}{\omega\beta}\HH_1\biggr)\bs_B \biggr]\rmdd\bx
&=&\int_\setdDD\bp_A^\dagger\biggl(\frac{2}{\omega\beta}\HH_1\biggr)^\dagger\bJ\bp_B n_3\rmd\bxh.
\end{eqnarray}
By substituting the expressions for $\bp$, $\bs$, $\bN$ and $\bJ$ (equations \ref{eqdecom4}, \ref{eq4.3NK} and \ref{eq4.3J}), using equations (\ref{eq90c}) and (\ref{eq90d}), we obtain
\begin{eqnarray}
&&-\int_\setD\frac{2}{\omega\beta}\bigl((\HH_1\s_A^+)\p_B^- - (\HH_1\s_A^-)\p_B^+ + \p_A^+(\HH_1\s_B^-) -\p_A^-(\HH_1\s_B^+) \bigr)\rmdd\bx\nonumber\\
&&\hspace{6cm}=-\int_\setdDD\frac{2}{\omega\beta}\bigl((\HH_1\p_A^+)\p_B^- - (\HH_1\p_A^-)\p_B^+\bigr) n_3\rmd\bxh,\label{eq66}\\
&&\int_\setD\frac{2}{\omega\beta}\bigl((\HH_1\s_A^+)^*\p_B^+ - (\HH_1\s_A^-)^*\p_B^- + \p_A^{+*}(\HH_1\s_B^+) -\p_A^{-*}(\HH_1\s_B^-) \bigr)\rmdd\bx\nonumber\\
&&\hspace{6cm}=\int_\setdDD\frac{2}{\omega\beta}\bigl((\HH_1\p_A^+)^*\p_B^+ - (\HH_1\p_A^-)^*\p_B^-\bigr) n_3\rmd\bxh.\label{eq67}
\end{eqnarray}
We aim to remove the operator $\HH_1$ from these equations.
From equation  (\ref{eqAAA26}) and (\ref{eqB19qf}) we obtain
\begin{eqnarray}
\partial_3\p^+&=&+\i{\cal H}_1\p^+ - \half({\cal L}_2^{-1}\partial_3{\cal L}_2)(\p^+ - \p^-)+\s^+,\label{eqAAA26b}\\
\partial_3\p^-&=&-\i{\cal H}_1\p^- + \half({\cal L}_2^{-1}\partial_3{\cal L}_2)(\p^+ - \p^-)+\s^-,\label{eqAAA26z}
\end{eqnarray}
with ${\cal L}_2$ defined in equation (\ref{eqB20qf}).
Assuming that in state $A$ the derivatives in the $x_3$-direction of the parameters $\alpha$ and $\beta$ at ${\setdDD}$ vanish and there are no sources at $\setdDD$, 
we find from equations (\ref{eqAAA26b}) and (\ref{eqAAA26z})
\begin{eqnarray}
&&\partial_3\p_A^\pm=\pm\i{\cal H}_1\p_A^\pm\quad\mbox{at} \,\,\,{\setdDD}.\label{eqAA46}
\end{eqnarray}
Below we use this to remove  $\HH_1$ from the right-hand sides of equations (\ref{eq66}) and (\ref{eq67}). Next, we aim to remove  $\HH_1$ from the left-hand sides of these 
equations.
From $\bs=\bL^{-1}\bd$ (equation \ref{eqdecom3}) and equations (\ref{eq2.2q}), (\ref{eqLi}), (\ref{eqdecom4}), (\ref{eqB19qf}) and (\ref{eqB20qf}), we find
\begin{eqnarray}
&&\s^\pm= \pm \half\Bigl(\frac{1}{\omega\beta}{\cal H}_1\Bigr)^{-1}B_0+ \half C_3,
\end{eqnarray}
or
\begin{eqnarray}
&&\pm\frac{2}{\omega\beta}{\cal H}_1\s^\pm= B_0\pm\frac{1}{\omega\beta}{\cal H}_1C_3.
\end{eqnarray}
We define new decomposed sources $B_0^+$ and $B_0^-$, according to
\begin{eqnarray}
&&B_0^\pm=B_0\pm\frac{1}{\omega\beta}{\cal H}_1C_3=\pm\frac{2}{\omega\beta}{\cal H}_1\s^\pm.\label{eqA33b}
\end{eqnarray}
Using equations (\ref{eqAA46}) and (\ref{eqA33b}) in the right- and left-hand sides of equations (\ref{eq66}) and (\ref{eq67}), we obtain
\begin{eqnarray}
\int_\setD\bigl(-B_{0,A}^+\p_B^- - B_{0,A}^-\p_B^+  + \p_A^+B_{0,B}^- + \p_A^-B_{0,B}^+\bigr)\rmdd\bx
&=&\int_\setdDD\frac{-2}{\i\omega\beta}\bigl((\partial_3\p_A^+)\p_B^- + (\partial_3\p_A^-)\p_B^+\bigr)n_3\rmd\bxh,\nonumber\\
&&\label{eq94}\\
\int_\setD\bigl(B_{0,A}^{+*}\p_B^+ + B_{0,A}^{-*}\p_B^-  + \p_A^{+*}B_{0,B}^+ + \p_A^{-*}B_{0,B}^-\bigr)\rmdd\bx
&=&\int_\setdDD\frac{-2}{\i\omega\beta}\bigl((\partial_3\p_A^+)^*\p_B^+ + (\partial_3\p_A^-)^*\p_B^-\bigr)n_3\rmd\bxh.\nonumber\\
&&\label{eq95}
\end{eqnarray}
Equations (\ref{eq94}) and (\ref{eq95}) are reciprocity theorems of the convolution type and correlation type, respectively, for field-normalised one-way wave fields.
These theorems are modifications of previously obtained results
%$^{43,44}$ 
\citep{Berkhout89GEO, Wapenaar89GEO}.
The main modification is that we applied decomposition at both sides of the equations instead of at the right-hand sides only. 
Moreover, in the present derivation the condition for the validity of equation (\ref{eqAA46}) is only imposed for state $A$.
In the next section we use equations (\ref{eq94}) and (\ref{eq95}) to derive representation theorems for field-normalised one-way wave fields and we  indicate applications.

\section{Field-normalised representation theorems}\label{sec5}

\subsection{Green's functions}

Representation theorems are obtained by substituting Green's functions in reciprocity theorems. Our aim is to introduce one-way Green's functions, to be used in 
the reciprocity theorems for field-normalised one-way wave fields (equations \ref{eq94} and \ref{eq95}).
First, we introduce the full Green's function $G(\bx,\bxA,\omega)$ as a solution of the unified wave equation (\ref{eqwe}) 
for a unit monopole point source at $\bxA$, with $B(\bx,\omega)= \delta(\bx-\bxA)$ and $C_j(\bx,\omega)=0$. Hence,
\begin{eqnarray}\label{eq6}
&&\beta\partial_j\Bigl(\frac{1}{\beta}\partial_j G\Bigr) +k^2G=\i\omega\beta \delta(\bx-\bxA).
\end{eqnarray}
As boundary condition we impose the radiation condition (i.e., outward propagating waves at infinity).
Next, we introduce one-way  Green's function as  solutions of the coupled one-way equations (\ref{eqAAA26b}) and (\ref{eqAAA26z}) for a unit monopole point source at $\bxA$.
Hence, we choose again $B(\bx,\omega)= \delta(\bx-\bxA)$ and $C_j(\bx,\omega)=0$. 
Using equations (\ref{eqA33b}) and  (\ref{eq2.5}), we define decomposed sources as $B_0^\pm=B^\pm=B=\pm 2 {\cal L}_2\s^\pm$, with ${\cal L}_2$ defined in equation (\ref{eqB20qf}), or
\begin{eqnarray}
&&\s^\pm(\bx,\omega)=\pm\half{\cal L}_2^{-1}B^\pm(\bx,\omega)=\pm\half{\cal L}_2^{-1}B(\bx,\omega)=\pm\half{\cal L}_2^{-1}\delta(\bx-\bxA).
\end{eqnarray}
We consider two sets of one-way Green's functions. 
For the first set we choose a point source $\s^+(\bx,\omega) =\half{\cal L}_2^{-1}B^+(\bx,\omega)$, with $B^+(\bx,\omega)=\delta(\bx-\bxA)$, 
which emits waves from $\bxA$ in the positive $x_3$-direction, 
and we set $\s^-(\bx,\omega) $ equal to zero.
Hence, for this first set, one-way equations (\ref{eqAAA26b}) and (\ref{eqAAA26z}) become
\begin{eqnarray}
\partial_3G^{+,+}&=&+\i{\cal H}_1G^{+,+} - \half({\cal L}_2^{-1}\partial_3{\cal L}_2)(G^{+,+} - G^{-,+})+\half{\cal L}_2^{-1}\delta(\bx-\bxA),\label{eqAAA26G1}\\
\partial_3G^{-,+}&=&-\i{\cal H}_1G^{-,+} + \half({\cal L}_2^{-1}\partial_3{\cal L}_2)(G^{+,+} - G^{-,+}).\label{eqAAA26G2}
\end{eqnarray}
Here $G^{\pm,+}$ stands for $G^{\pm,+}(\bx,\bxA,\omega)$. The second superscript ($+$) indicates that the source at $\bxA$ emits waves in the positive $x_3$-direction.
The first superscript  ($\pm$)  denotes the propagation direction at $\bx$.
For the second set of one-way Green's functions we choose a point source 
$\s^-(\bx,\omega) =-\half{\cal L}_2^{-1}B^-(\bx,\omega)$, with $B^-(\bx,\omega)=\delta(\bx-\bxA)$, which emits waves from $\bxA$ in the negative $x_3$-direction, 
and we set $\s^+(\bx,\omega) $ equal to zero.
Hence, for this second set, one-way equations (\ref{eqAAA26b}) and (\ref{eqAAA26z}) become
\begin{eqnarray}
\partial_3G^{+,-}&=&+\i{\cal H}_1G^{+,-} - \half({\cal L}_2^{-1}\partial_3{\cal L}_2)(G^{+,-} - G^{-,-}),\label{eqAAA26G3}\\
\partial_3G^{-,-}&=&-\i{\cal H}_1G^{-,-} + \half({\cal L}_2^{-1}\partial_3{\cal L}_2)(G^{+,-} - G^{-,-})-\half{\cal L}_2^{-1}\delta(\bx-\bxA).\label{eqAAA26G4}
\end{eqnarray}
Here $G^{\pm,-}$ stands for $G^{\pm,-}(\bx,\bxA,\omega)$, with the second superscript ($-$) indicating that the source at $\bxA$ emits waves in the negative $x_3$-direction.
Like for the full Green's function $G(\bx,\bxA,\omega)$, we impose radiation conditions for both sets of one-way Green's functions.

To find a relation between the full Green's function and the one-way Green's functions, 
we evaluate $\beta\partial_3\frac{1}{\beta}\partial_3(G^{+,+}+G^{-,+}+G^{+,-}+G^{-,-})$ using 
equations (\ref{eqAAA26G1}) $-$ (\ref{eqAAA26G4}), 
(\ref{eq39}), (\ref{eqB18qf}) and (\ref{eqB20qf}). This gives equation (\ref{eq6}), with $G$ replaced by $G^{+,+}+G^{-,+}+G^{+,-}+G^{-,-}$.
Since the full Green's function and the one-way Green's functions obey the same radiation conditions, we thus find
\begin{eqnarray}
&&G=G^{+,+}+G^{-,+}+G^{+,-}+G^{-,-}.\label{eq106}
\end{eqnarray}
This very simple relation is a consequence of the field-normalised decomposition, introduced in section \ref{sec3c}.

\subsection{Source-receiver reciprocity}

\begin{figure}
\vspace{0cm}
\centerline{\epsfysize=7.5 cm \epsfbox{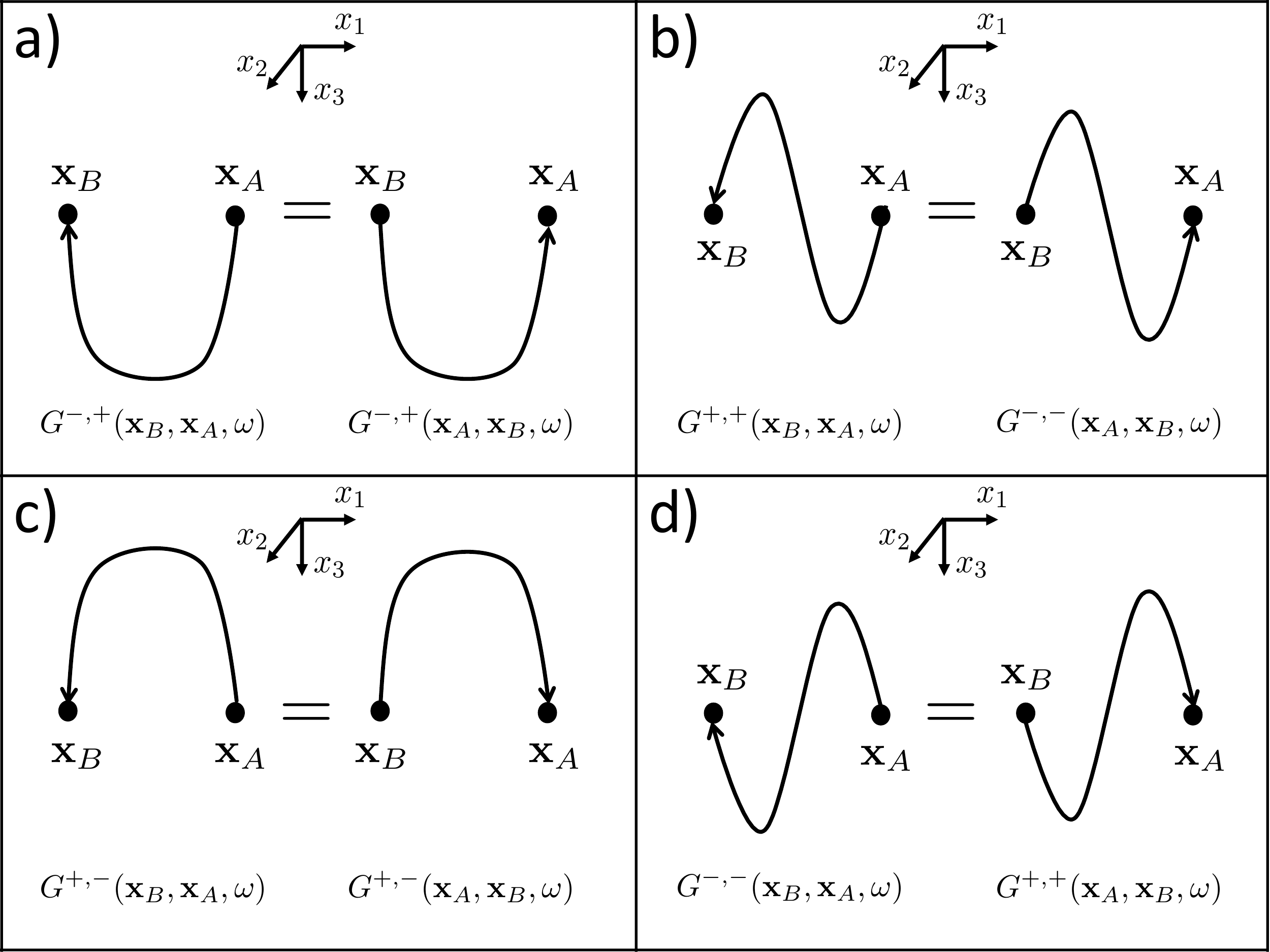}}
\caption{\footnotesize Visualisation of the source-receiver reciprocity relations for the field-normalised one-way Green's functions, 
formulated by equations (\ref{eqA50}) $-$ (\ref{eqA53}). The ``rays'' in this and subsequent figures are strong simplifications of the complete one-way wave fields, 
which include primary and multiple scattering. 
}\label{Fig2}
\end{figure}

We derive source-receiver reciprocity relations for the field-normalised one-way Green's functions introduced in the previous section.
To this end we make use of the reciprocity theorem of the convolution type for field-normalised one-way wave fields,
equation (\ref{eq94}). This theorem was derived for the configuration of Figure \ref{Fig1}, assuming that
in domain $\setD$, the parameters $\alpha$ and $\beta$ are the same in the two states (see section \ref{sec3d}).
Outside $\setD$ these parameters may be different in the two states. For the Green's state we choose the parameters \rev{for $x_3\le x_{3,0}$ and for $x_3\ge x_{3,1}$}
 independent of the $x_3$-coordinate, according to $\alpha(\bxh)$ and $\beta(\bxh)$. 
Hence, if we let the Green's state (with a point source at $\bxA$ in $\setD$) take the role of state $A$, then the condition for the validity of equation (\ref{eqAA46}) is fulfilled.
Moreover, the Green's functions are purely outward propagating at $\setdDD$ (because outside $\setD$ no scattering occurs along the $x_3$-coordinate). 
Hence, $G^{+,\pm}(\bx,\bxA,\omega)=0$ at $\setdD_0$ and $G^{-,\pm}(\bx,\bxA,\omega)=0$ at $\setdD_1$.
We let a second Green's state (with a point source at $\bxB$ in $\setD$ and the same parameters $\alpha$ and $\beta$ \rev{as in state $A$}, inside as well as outside $\setD$) take the role of state $B$. 
Hence, $G^{+,\pm}(\bx,\bxB,\omega)=0$ at $\setdD_0$ and $G^{-,\pm}(\bx,\bxB,\omega)=0$ at $\setdD_1$.
With only outward propagating waves at $\setdDD$, the surface integral on the right-hand side of equation (\ref{eq94}) vanishes.
Hence, taking into account that $B_0^\pm=B^\pm$ (since $C_j=0$), equation (\ref{eq94}) simplifies to
\begin{eqnarray}\label{eq4.1g}
&&\int_{\setD}\bigl (-B_A^+\p_B^- - B_A^-\p_B^+  + \p_A^+B_B^- + \p_A^-B_B^+\bigr)\rmdd\bx=0.
\end{eqnarray}
First, we consider sources emitting waves in the positive $x_3$-direction in both Green's  states, hence $B_A^+=\delta(\bx-\bx_A)$, $B_A^-=0$, $\p_A^\pm=G^{\pm,+}(\bx,\bx_A,\omega)$, $B_B^+=\delta(\bx-\bx_B)$, $B_B^-=0$ and $\p_B^\pm=G^{\pm,+}(\bx,\bx_B,\omega)$. Substituting this into equation (\ref{eq4.1g}) yields
\begin{eqnarray}
&&G^{-,+}(\bx_B,\bx_A,\omega)=G^{-,+}(\bx_A,\bx_B,\omega),\label{eqA50}
\end{eqnarray}
see Figure \ref{Fig2}(a).
Next, we replace the source in state $B$ by one emitting waves in the negative $x_3$-direction, hence
$B_B^+=0$, $B_B^-=\delta(\bx-\bx_B)$ and $\p_B^\pm=G^{\pm,-}(\bx,\bx_B,\omega)$.
This gives
\begin{eqnarray}
&&G^{+,+}(\bx_B,\bx_A,\omega)=G^{-,-}(\bx_A,\bx_B,\omega),\label{eqA51}
\end{eqnarray}
see Figure \ref{Fig2}(b).
By replacing also the source in state $A$ by one emitting waves in the negative $x_3$-direction, according to
 $B_A^+=0$, $B_A^-=\delta(\bx-\bx_A)$ and $\p_A^\pm=G^{\pm,-}(\bx,\bx_A,\omega)$, 
we obtain
\begin{eqnarray}
&&G^{+,-}(\bx_B,\bx_A,\omega)=G^{+,-}(\bx_A,\bx_B,\omega),\label{eqA52}
\end{eqnarray}
see Figure \ref{Fig2}(c). Finally, changing the source in state $B$ back to the one emitting waves in the positive $x_3$-direction yields
\begin{eqnarray}
&&G^{-,-}(\bx_B,\bx_A,\omega)=G^{+,+}(\bx_A,\bx_B,\omega),\label{eqA53}
\end{eqnarray}
see Figure \ref{Fig2}(d). 

Source-receiver reciprocity relations similar to equations  (\ref{eqA50}) $-$ (\ref{eqA53}) 
were previously derived for flux-normalised one-way Green's functions
%$^{17}$ 
\citep{Wapenaar96GJI2}, 
except that two of those relations involve a change of sign when interchanging the source and the receiver. 
The absence of sign changes in equations  (\ref{eqA50}) $-$ (\ref{eqA53}) 
is due to the definition of $B_0^\pm$ in equation (\ref{eqA33b}).
Moreover, unlike the flux-normalised reciprocity relations, the field-normalised source-receiver reciprocity relations of equations  
(\ref{eqA50}) $-$ (\ref{eqA53}) have a very straightforward relation with the
well-known source-receiver reciprocity relation for the full Green's function.
By separately summing the left- and right-hand sides of equations (\ref{eqA50}) $-$ (\ref{eqA53}) and using equation (\ref{eq106}),
we simply obtain  
\begin{eqnarray}
&&G(\bx_B,\bx_A,\omega)=G(\bx_A,\bx_B,\omega).\label{eqA54}
\end{eqnarray}

\subsection{Kirchhoff-Helmholtz integrals for forward propagation}

\begin{figure}
\vspace{0cm}
\centerline{\epsfysize=8 cm \epsfbox{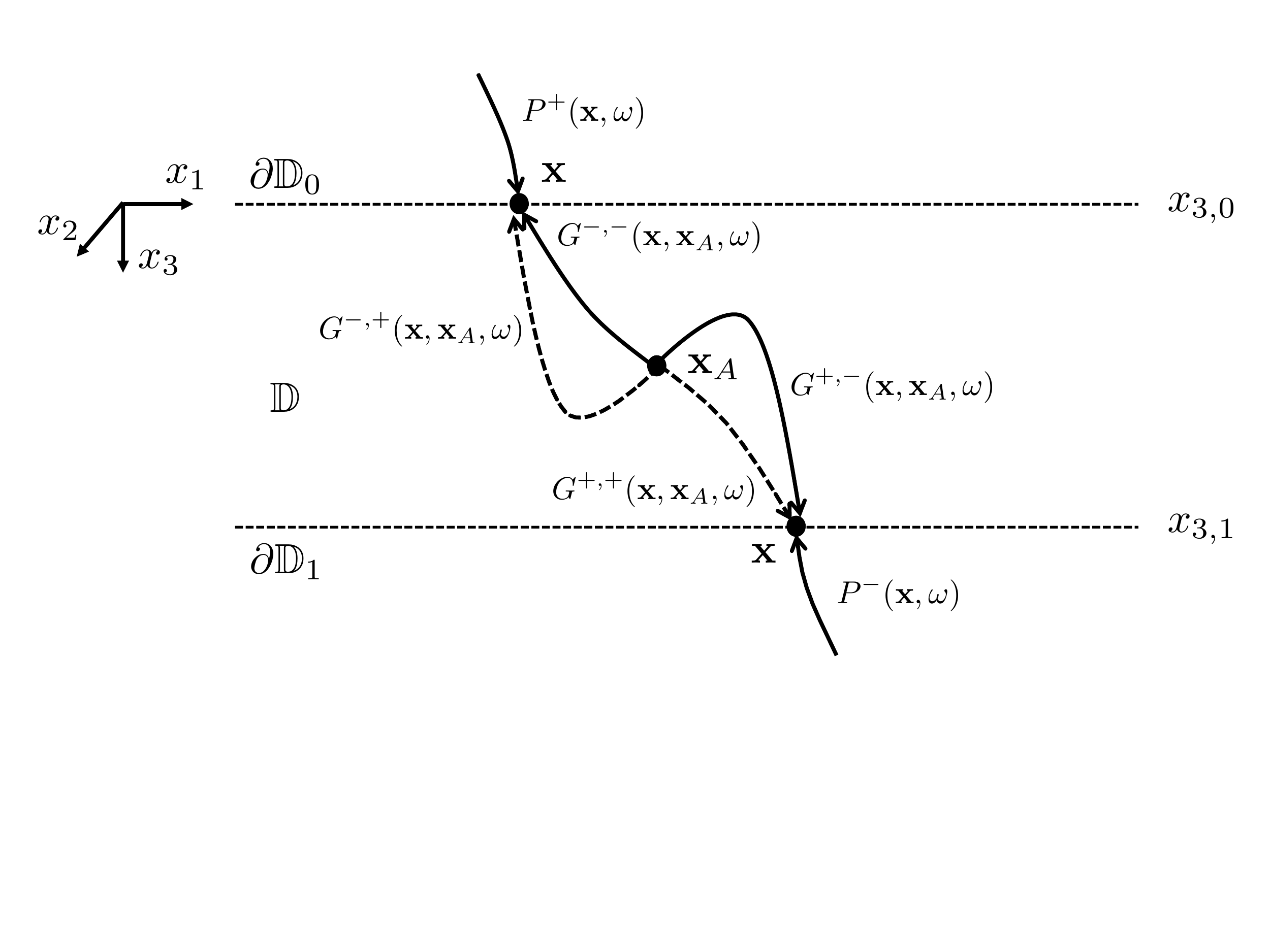}}
\vspace{-1.8cm}
\caption{\footnotesize Visualisation of the different terms in the field-normalised one-way Kirchhoff-Helmholtz integral for forward propagation, 
formulated by equation (\ref{eq4.1compqGcc}). 
The solid Green's functions contribute to $\p^+(\bxA,\omega)$, the dashed Green's functions to $\p^-(\bxA,\omega)$.
}\label{Fig3}
\end{figure}

We derive Kirchhoff-Helmholtz integrals of the convolution type for field-normalised one-way wave fields.
For state $B$ we consider the decomposed actual field, with  sources only outside $\setD$, hence, 
$B_{0,B}^\pm=0$ in $\setD$ and $\p_B^\pm=\p^\pm(\bx,\omega)$. 
The  parameters $\alpha$ and $\beta$  are the actual parameters inside as well as outside $\setD$.
For state $A$ we choose the Green's state with a unit point source at $\bxA$ in $\setD$.
The parameters $\alpha$ and $\beta$ in $\setD$ are the same as those in state $B$, but \rev{for $x_3\le x_{3,0}$ and for $x_3\ge x_{3,1}$} they are chosen independent of the $x_3$-coordinate.
Hence, the condition for the validity of equation (\ref{eqAA46}) is fulfilled.
First, we consider a source in state $A$ which emits waves in the positive $x_3$-direction, hence
 $B_A^+=\delta(\bx-\bxA)$, $B_A^-=0$ and $P_A^\pm=G^{\pm,+}(\bx,\bxA,\omega)$.
Substituting all this into equation (\ref{eq94}) (with $B_{0,A}^\pm=B_A^\pm$)  gives
\begin{eqnarray}
&&\p^-(\bxA,\omega)=
\int_\setdDD\frac{2}{\i\omega\beta(\bx)}\bigl((\partial_3G^{+,+} (\bx,\bxA,\omega))\p^-(\bx,\omega) 
+ (\partial_3G^{-,+} (\bx,\bxA,\omega))\p^+ (\bx,\omega)\bigr)n_3\rmd\bxh.\nonumber\\
&&\label{eq4.1compqGaa}
\end{eqnarray}
Next, we replace the  source in state $A$ by one which emits waves in the negative $x_3$-direction, hence
$B_A^+=0$, $B_A^-=\delta(\bx-\bxA)$ and $P_A^\pm=G^{\pm,-}(\bx,\bxA,\omega)$.
Equation (\ref{eq94}) thus gives
\begin{eqnarray}
&&\p^+(\bxA,\omega)=
\int_\setdDD\frac{2}{\i\omega\beta(\bx)}\bigl((\partial_3G^{+,-} (\bx,\bxA,\omega))\p^-(\bx,\omega) 
+ (\partial_3G^{-,-} (\bx,\bxA,\omega))\p^+ (\bx,\omega)\bigr)n_3\rmd\bxh.\nonumber\\
&&\label{eq4.1compqGbb}
\end{eqnarray}
Recall that $\setdDD$ consists of $\setdD_0$ (with $n_3=-1$) and $\setdD_1$ (with $n_3=+1$), see Figure \ref{Fig1}.
Since  $G^{+,\pm}(\bx,\bxA,\omega)=0$ at $\setdD_0$ and $G^{-,\pm}(\bx,\bxA,\omega)=0$ at $\setdD_1$
 (because outside $\setD$ no scattering occurs along the $x_3$-coordinate in state $A$), the first term under the integral in 
equations (\ref{eq4.1compqGaa}) and  (\ref{eq4.1compqGbb}) gives a contribution only at $\setdD_1$ and the second term only at $\setdD_0$.
Hence,
\begin{eqnarray}
\p^\pm(\bxA,\omega)&=&
\int_{\setdD_0}\frac{-2}{\i\omega\beta(\bx)}(\partial_3G^{-,\mp} (\bx,\bxA,\omega))\p^+(\bx,\omega) \rmd\bxh\nonumber\\
&+&\int_{\setdD_1}\frac{2}{\i\omega\beta(\bx)} (\partial_3G^{+,\mp} (\bx,\bxA,\omega))\p^- (\bx,\omega)\bigr)\rmd\bxh.\label{eq4.1compqGcc}
\end{eqnarray}
Note that there is no contribution from $\p^- (\bx,\omega)$ at $\setdD_0$ nor from $\p^+ (\bx,\omega)$ at $\setdD_1$,
see Figure \ref{Fig3}. 

We conclude this section by considering a special case. Suppose the source of the actual field (state $B$) is located at $\bxB$ in the half-space $x_3<x_{3,0}$.
Then, by taking $x_{3,1}\to\infty$, the field $\p^-$ at $\setdD_1$ vanishes. This leaves the single-sided representation
\begin{eqnarray}
\p^\pm(\bxA,\bxB,\omega)&=&
\int_{\setdD_0}\frac{-2}{\i\omega\beta(\bx)}(\partial_3G^{-,\mp} (\bx,\bxA,\omega))\p^+(\bx,\bxB,\omega) \rmd\bxh.\label{eq4.1compqGee}
\end{eqnarray}
Note that we included the source coordinate vector $\bxB$ in the argument list of $\p^\pm(\bxA,\bxB,\omega)$.
This representation is an extension of a previously derived result
%$^{43}$
\citep{Berkhout89GEO}, 
in which the fields were decomposed at $\setdD_0$ but not at $\bxA$.
It describes forward propagation of the one-way field $\p^+(\bx,\bxB,\omega)$ from the surface $\setdD_0$ to $\bxA$ (with $\bxA$ and $\bxB$ defined at opposite sides of $\setdD_0$).
In the following two sections we discuss representations for backward propagation of one-way wave fields.

\subsection{Kirchhoff-Helmholtz integrals for backward propagation (double-sided)}\label{sec4d}

\begin{figure}
\vspace{0cm}
\centerline{\epsfysize=8 cm \epsfbox{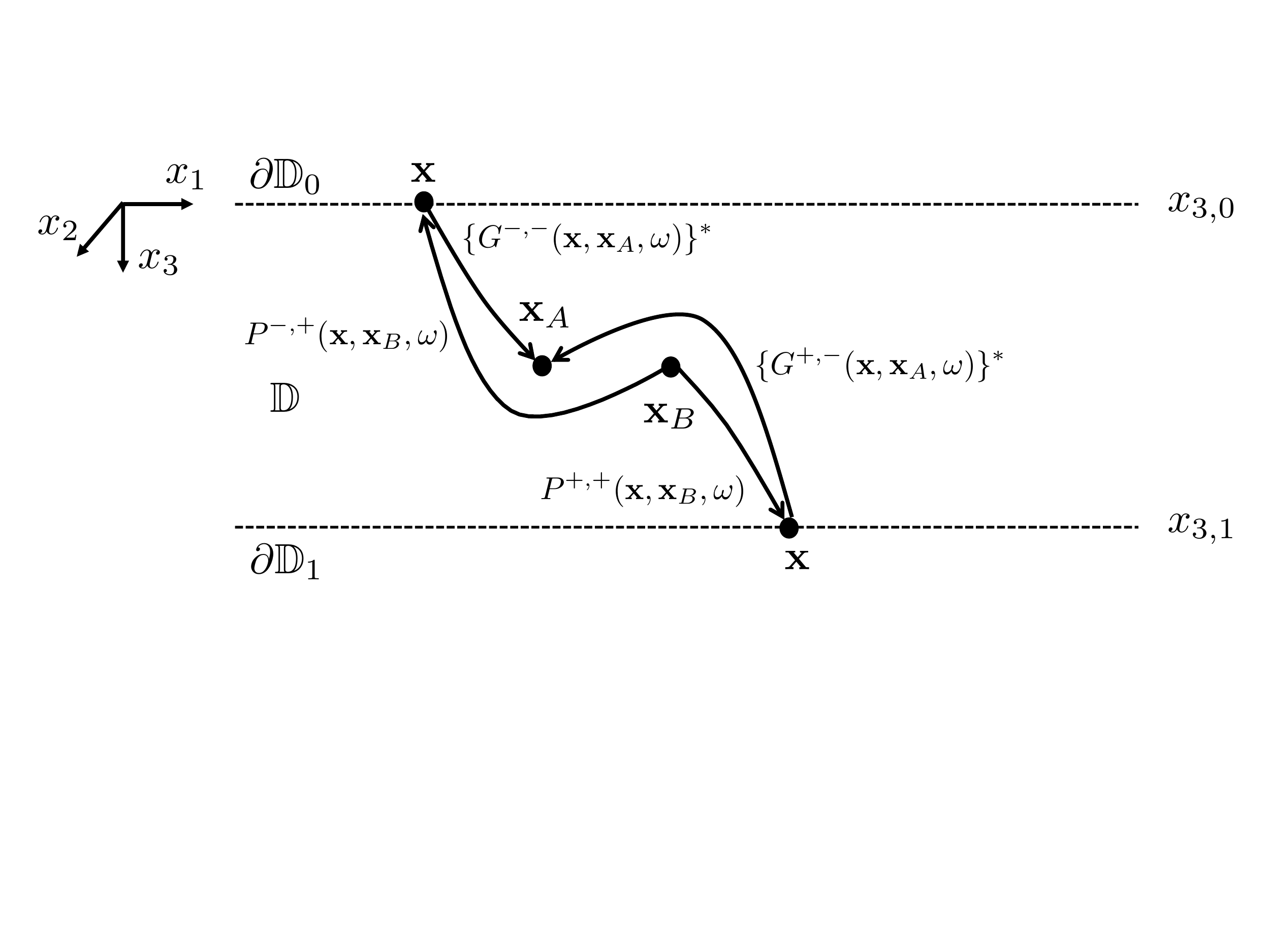}}
\vspace{-2.8cm}
\caption{\footnotesize Visualisation of the different terms in the field-normalised one-way Kirchhoff-Helmholtz integral for backward propagation, 
formulated by equation (\ref{eq104c-altb}). 
}\label{Fig4a}
\end{figure}

We derive Kirchhoff-Helmholtz integrals of the correlation type for field-normalised one-way wave fields.
For state $B$ we consider the decomposed actual field, with a point source at $\bxB$ and  source spectrum $\ss(\omega)$. 
The  parameters $\alpha$ and $\beta$ are the actual parameters inside as well as outside $\setD$.
For state $A$ we choose the Green's state with a unit point source at $\bxA$ in $\setD$.
The parameters $\alpha$ and $\beta$ in $\setD$ are the same as those in state $B$, but \rev{for $x_3\le x_{3,0}$ and for $x_3\ge x_{3,1}$} they are chosen independent of the $x_3$-coordinate.
Hence, the condition for the validity of equation (\ref{eqAA46}) is fulfilled.
First, we consider sources emitting waves in the positive $x_3$-direction in both states, hence 
$B_A^+=\delta(\bx-\bx_A)$, $B_A^-=0$, $\p_A^\pm=G^{\pm,+}(\bx,\bx_A,\omega)$, 
$B_B^+=\delta(\bx-\bx_B)\ss(\omega)$, $B_B^-=0$ and $\p_B^\pm=P^{\pm,+}(\bx,\bx_B,\omega)$.
Substituting this into equation (\ref{eq95}) 
(with $B_{0,A}^\pm=B_A^\pm$ and  $B_{0,B}^\pm=B_B^\pm$) 
gives
\begin{eqnarray}\label{eq104a-alt}
&&P^{+,+}(\bx_A,\bx_B,\omega)+\chi(\bxB)\{G^{+,+}(\bx_B,\bx_A,\omega)\}^*\ss(\omega)=\\
&&\int_\setdDD\frac{-2}{\i\omega\beta(\bx)}\bigl(\{\partial_3G^{+,+} (\bx,\bxA,\omega)\}^*P^{+,+}(\bx,\bxB,\omega) 
+\{\partial_3G^{-,+} (\bx,\bxA,\omega)\}^*P^{-,+}(\bx,\bxB,\omega)\bigr)n_3\rmd\bxh\nonumber,
\end{eqnarray}
where $\chi$ is the characteristic function of the domain $\setD$. It is defined as
\begin{eqnarray}\label{eqC3.2D}
&&\chi(\bxB)=
\begin{cases}
1,   &\text{for } \bxB\text{ in }{\setD}, \\
\half, &\text{for } \bxB\text{ on }{\setdDD},\\
 0,  &\text{for } \bxB\text{ outside }{\setD}.
 \end{cases}
\end{eqnarray}
Since  $G^{+,+}(\bx,\bxA,\omega)=0$ at $\setdD_0$ and $G^{-,+}(\bx,\bxA,\omega)=0$ at $\setdD_1$
 (because outside $\setD$ no scattering occurs along the $x_3$-coordinate in state $A$), the first term under the integral in 
equation (\ref{eq104a-alt})  gives a contribution only at $\setdD_1$ and the second term only at $\setdD_0$.
Hence,
\begin{eqnarray}\label{eq104a-altb}
&&P^{+,+}(\bx_A,\bx_B,\omega)+\chi(\bxB)\{G^{+,+}(\bx_B,\bx_A,\omega)\}^*\ss(\omega)=\\
&&\int_{\setdD_0}\frac{2}{\i\omega\beta(\bx)}\{\partial_3G^{-,+} (\bx,\bxA,\omega)\}^*P^{-,+}(\bx,\bxB,\omega)\rmd\bxh\nonumber\\
&&-\int_{\setdD_1}\frac{2}{\i\omega\beta(\bx)}\{\partial_3G^{+,+} (\bx,\bxA,\omega)\}^*P^{+,+}(\bx,\bxB,\omega)\rmd\bxh\nonumber.
\end{eqnarray}
Next, we replace the source in state $B$ by one emitting waves in the negative $x_3$-direction, hence
$B_B^+=0$, $B_B^-=\delta(\bx-\bx_B)\ss(\omega)$ and $\p_B^\pm=P^{\pm,-}(\bx,\bx_B,\omega)$.
This gives
\begin{eqnarray}\label{eq104b-altb}
&&P^{+,-}(\bx_A,\bx_B,\omega)+\chi(\bxB)\{G^{-,+}(\bx_B,\bx_A,\omega)\}^*\ss(\omega)=\\
&&\int_{\setdD_0}\frac{2}{\i\omega\beta(\bx)}\{\partial_3G^{-,+} (\bx,\bxA,\omega)\}^*P^{-,-}(\bx,\bxB,\omega)\rmd\bxh\nonumber\\
&&-\int_{\setdD_1}\frac{2}{\i\omega\beta(\bx)}\{\partial_3G^{+,+} (\bx,\bxA,\omega)\}^*P^{+,-}(\bx,\bxB,\omega) \rmd\bxh\nonumber.
\end{eqnarray}
By replacing also the source in state $A$ by one emitting waves in the negative $x_3$-direction, according to
 $B_A^+=0$, $B_A^-=\delta(\bx-\bx_A)$,  $\p_A^\pm=G^{\pm,-}(\bx,\bx_A,\omega)$, 
we obtain
\begin{eqnarray}\label{eq104d-altb}
&&P^{-,-}(\bx_A,\bx_B,\omega)+\chi(\bxB)\{G^{-,-}(\bx_B,\bx_A,\omega)\}^*\ss(\omega)=\\
&&\int_{\setdD_0}\frac{2}{\i\omega\beta(\bx)}\{\partial_3G^{-,-} (\bx,\bxA,\omega)\}^*P^{-,-}(\bx,\bxB,\omega)\rmd\bxh\nonumber\\
&&-\int_{\setdD_1}\frac{2}{\i\omega\beta(\bx)}\{\partial_3G^{+,-} (\bx,\bxA,\omega)\}^*P^{+,-}(\bx,\bxB,\omega) \rmd\bxh\nonumber.
\end{eqnarray}
Finally, changing the source in state $B$ back to the one emitting waves in the positive $x_3$-direction yields
\begin{eqnarray}\label{eq104c-altb}
&&P^{-,+}(\bx_A,\bx_B,\omega)+\chi(\bxB)\{G^{+,-}(\bx_B,\bx_A,\omega)\}^*\ss(\omega)=\\
&&\int_{\setdD_0}\frac{2}{\i\omega\beta(\bx)}\{\partial_3G^{-,-} (\bx,\bxA,\omega)\}^*P^{-,+}(\bx,\bxB,\omega)\rmd\bxh\nonumber\\
&&-\int_{\setdD_1}\frac{2}{\i\omega\beta(\bx)}\{\partial_3G^{+,-} (\bx,\bxA,\omega)\}^*P^{+,+}(\bx,\bxB,\omega)\rmd\bxh\nonumber.
\end{eqnarray}
Equation (\ref{eq104d-altb}) is an extension of a previously derived result
%$^{44}$  
\citep{Wapenaar89GEO}, 
in which the fields were decomposed at $\setdDD$ but not at $\bxA$ and $\bxB$.
Equations (\ref{eq104a-altb}), (\ref{eq104b-altb}) and (\ref{eq104c-altb}) are further variations. 
Equation (\ref{eq104c-altb}) is visualised in Figure \ref{Fig4a}.
Together,  these equations describe backward propagation of the one-way wave fields $P^{-,\pm} (\bx,\bxB,\omega)$ from $\setdD_0$ and $P^{+,\pm} (\bx,\bxB,\omega)$ 
from $\setdD_1$ to $\bxA$. 
Except for some special cases,  the integrals along $\setdD_1$ do not vanish by taking $x_{3,1}\to\infty$. 
Hence, unlike the forward propagation representation (\ref{eq4.1compqGcc}), the double-sided backward propagation representations (\ref{eq104a-altb}) $-$ (\ref{eq104c-altb})  
in general do not simplify to  single-sided representations. In the next section we discuss an alternative method to derive single-sided representations for backward propagation.

We conclude this section by considering a special case.
Suppose that 
in state $B$ the parameters $\alpha$ and $\beta$   are the same as in state $A$ not only in $\setD$ but also outside $\setD$.
Then $P^{\pm,\pm}(\bx,\bxB,\omega)=G^{\pm,\pm}(\bx,\bxB,\omega)\ss(\omega)$
for all $\bx$. Substituting this into representations (\ref{eq104a-altb}) $-$ (\ref{eq104c-altb}), 
summing the left- and right-hand sides of these representations separately and dividing both sides by $\ss(\omega)$,  
using equations (\ref{eq106}) and (\ref{eqA54}) and assuming that $\bxB$ is located in $\setD$, we obtain
\begin{eqnarray}\label{eq104acbd-altb}
G_{\rm h}(\bx_A,\bx_B,\omega)
&=&\int_{\setdD_0}\frac{2}{\i\omega\beta(\bx)}\{\partial_3G^- (\bx,\bxA,\omega)\}^*G^-(\bx,\bxB,\omega)\rmd\bxh\\
&-&\int_{\setdD_1}\frac{2}{\i\omega\beta(\bx)}\{\partial_3G^+ (\bx,\bxA,\omega)\}^*G^+(\bx,\bxB,\omega) \rmd\bxh\nonumber,
\end{eqnarray}
where the so-called homogeneous Green's function $G_{\rm h}(\bx_A,\bx_B,\omega)$ is defined as
\begin{eqnarray}\label{eq100}
&&G_{\rm h}(\bx_A,\bx_B,\omega)=G(\bx_A,\bx_B,\omega)+G^*(\bx_A,\bx_B,\omega)=2\Re \{G(\bx_A,\bx_B,\omega)\}
\end{eqnarray}
(with $\Re$ denoting the real part),
and where $G^\pm(\bx,\bxA,\omega)=G^{\pm,+}(\bx,\bxA,\omega)+G^{\pm,-}(\bx,\bxA,\omega)$ (and a similar expression for $G^\pm(\bx,\bxB,\omega)$).
Equation (\ref{eq104acbd-altb}) is akin to the well-known representation for the homogeneous Green's function
%$^{45,46}$ 
\citep{Porter70JOSA, Oristaglio89IP}, 
but with decomposed Green's functions under the integrals.
The simple relation between representations (\ref{eq104a-altb}) $-$ (\ref{eq104c-altb}) on the one hand and the homogeneous Green's function representation (\ref{eq104acbd-altb})  
on the other hand is   a consequence of the field-normalised decomposition, introduced in section \ref{sec3c}.

\subsection{Kirchhoff-Helmholtz integrals for backward propagation (single-sided)}\label{sec4e}

The complex-conjugated Green's functions $\{\partial_3G^{\pm,\pm}(\bx,\bxA,\omega)\}^*$ under the integrals in equations (\ref{eq104a-altb}) $-$ (\ref{eq104c-altb}) 
can be seen as focusing functions,  which focus the wave fields $P^{\pm,\pm}(\bx,\bxB,\omega)$  onto a focal point  $\bxA$.
However, this focusing process requires that these wave fields are available at two boundaries $\setdD_0$ and $\setdD_1$, enclosing the focal point $\bxA$.
Here we discuss single-sided field-normalised 
focusing functions $f_1^\pm(\bx,\bxA,\omega)$ and we use these in modifications of reciprocity theorems (\ref{eq94}) and (\ref{eq95}) to derive
single-sided Kirchhoff-Helmholtz integrals for backward propagation.

\begin{figure}
\vspace{0cm}
\centerline{\epsfysize=8 cm \epsfbox{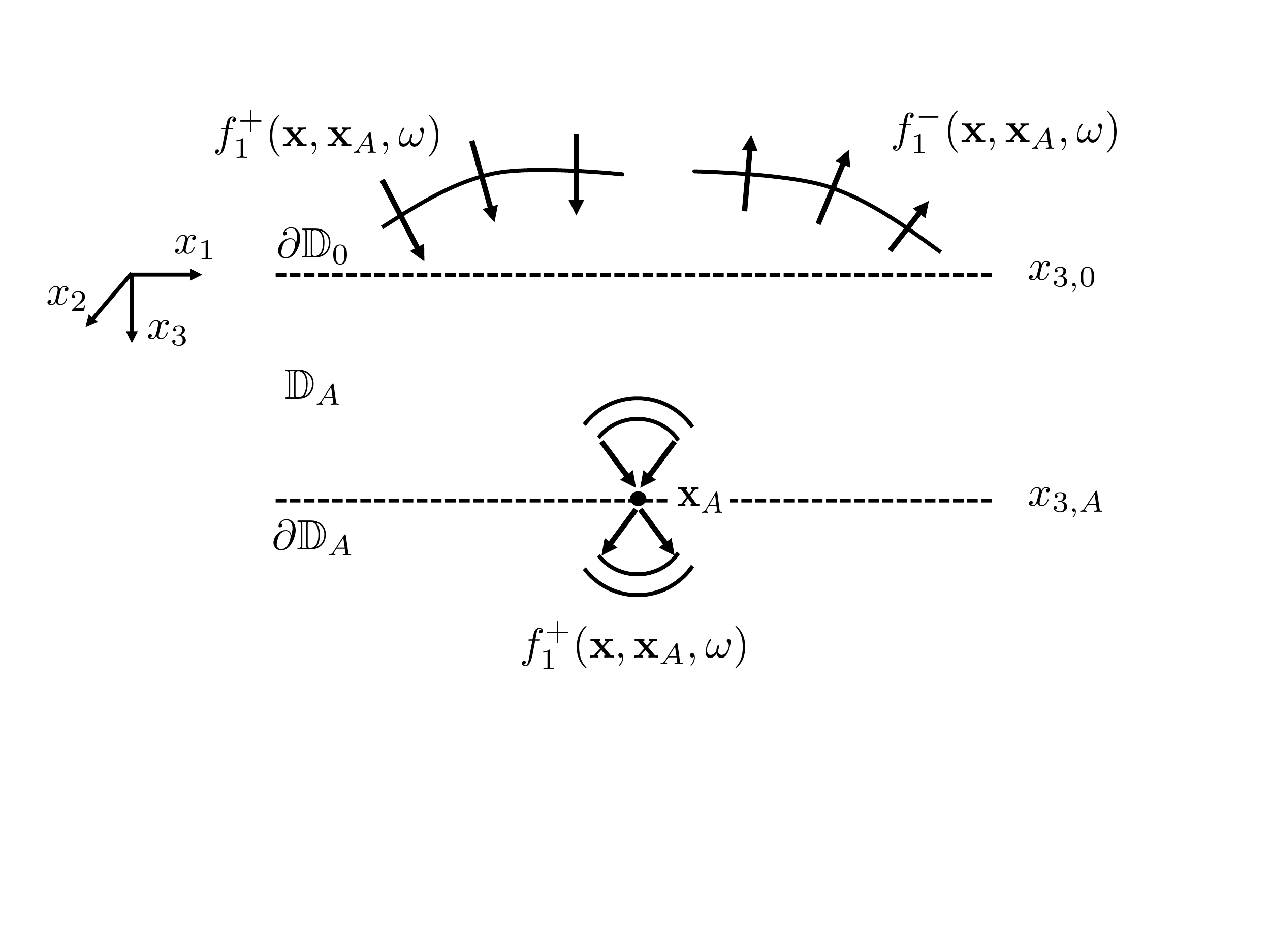}}
\vspace{-1.8cm}
\caption{\footnotesize 
Configuration for the derivation of the single-sided Kirchhoff-Helmholtz integrals for backward propagation.
}\label{Fig4}
\end{figure}

We start by defining a new domain $\setD_A$, enclosed by two  surfaces $\setdD_0$ and $\setdD_A$ perpendicular to the $x_3$-axis at $x_3=x_{3,0}$ and $x_3=x_{3,A}$, 
respectively, with $x_{3,A}> x_{3,0}$, see Figure \ref{Fig4}. Hence, $\setdD_A$ is chosen such that it contains the focal point $\bxA$.
The two surfaces $\setdD_0$ and $\setdD_A$ are together denoted by $\setdDA$. 
The focusing functions $f_1^\pm(\bx,\bxA,\omega)$, which will play the role of state $A$ in the reciprocity theorems, obey the one-way wave equations (\ref{eqAAA26b}) and 
(\ref{eqAAA26z}) (but without the source terms $\s^\pm$), with parameters $\alpha$ and $\beta$ in $\setD_A$ equal to those in the actual state $B$,
and independent of the $x_3$-coordinate \rev{for $x_3\le x_{3,0}$ and for $x_3\ge x_{3,A}$}. Hence, the condition for the validity of equation (\ref{eqAA46}) is fulfilled.
Analogous to equation (\ref{eq59}), the field-normalised focusing functions $f_1^\pm(\bx,\bxA,\omega)$ are related to the full focusing function $f_1(\bx,\bxA,\omega)$, according
to
\begin{eqnarray}\label{eq101}
&&f_1(\bx,\bxA,\omega)=f_1^+(\bx,\bxA,\omega)+f_1^-(\bx,\bxA,\omega).
\end{eqnarray}
The focusing function $f_1^+(\bx,\bxA,\omega)$ is incident to the domain $\setD_A$ from the half-space $x_3<x_{3,0}$ (see Figure \ref{Fig4}). It propagates and scatters in the
inhomogeneous domain $\setD_A$, focuses at $\bxA$ on surface $\setdD_A$ and continues as $f_1^+(\bx,\bxA,\omega)$ in the half-space $x_3>x_{3,A}$. 
The back-scattered field leaves $\setD_A$ via surface $\setdD_0$ and continues as $f_1^-(\bx,\bxA,\omega)$ in the half-space $x_3<x_{3,0}$.
The focusing conditions at the focal plane  $\setdD_A$ are
%$^{18}$ 
\citep{Wapenaar2014GEO}
\begin{eqnarray}
&&[\partial_3f_1^+(\bx,\bx_A,\omega)]_{x_3=x_{3,A}}=\half \i\omega\beta(\bx_A)\delta(\bxh-\bxha),\label{eqAA47}\\
&&[\partial_3f_1^-(\bx,\bx_A,\omega)]_{x_3=x_{3,A}}=0.\label{eqAA48}
\end{eqnarray}
Here $\bxha$ denotes the lateral coordinates of $\bxA$. The operators $\partial_3$ and the factor $\half \i\omega\beta(\bx_A)$ are 
not necessary to define the focusing conditions but are chosen for later convenience.
To avoid instability, evanescent waves are excluded from the focusing functions. This implies that
the delta function in equation (\ref{eqAA47}) should be interpreted as a spatially band-limited delta function. Note that the sifting property of the delta function, 
$h(\bxha)=\int_\setA\delta(\bxh-\bxha)h(\bxh)\rmd\bxh$, remains valid for a spatially band-limited delta function, assuming $h(\bxh)$ is also spatially band-limited.

We now derive single-sided Kirchhoff-Helmholtz integrals for backward propagation. 
We consider the reciprocity theorems for field-normalised one-way wave fields (equations \ref{eq94} and \ref{eq95}), with $\setD$ \rev{and $\setdDD$ replaced by $\setD_A$ and $\setdDA$,
respectively.}
For state $A$ we consider the focusing functions discussed above, hence, $B_A^+(\bx,\omega)=B_A^-(\bx,\omega)=0$ and $\p_A^\pm(\bx,\omega)=f_1^\pm(\bx,\bx_A,\omega)$.
For state $B$ we consider the decomposed actual field, with a point source at $\bxB$ in the half-space $x_3>x_{3,0}$  and source spectrum $\ss(\omega)$. 
The  parameters $\alpha$ and $\beta$ in state $B$ are the actual parameters inside as well as outside $\setdDA$.
First, we consider a source in state $B$ which emits waves in the positive $x_3$-direction, hence
$B_B^+(\bx,\omega)=\delta(\bx-\bx_B)\ss(\omega)$, $B_B^-(\bx,\omega)=0$ 
and $\p_B^\pm(\bx,\omega)=P^{\pm,+}(\bx,\bx_B,\omega)$. Substituting all this
into equations (\ref{eq94}) and (\ref{eq95}) (with $B_0^\pm=B^\pm$), 
using equations (\ref{eqAA47}) and  (\ref{eqAA48}) in the integrals along $\setdD_A$, gives
\begin{eqnarray}
&&P^{-,+}(\bx_A,\bx_B,\omega)+\chi_A(\bxB)f_1^-(\bx_B,\bx_A,\omega)\ss(\omega)\label{eqAA60-alt}\\
&&=\int_{\setdD_0}\frac{2}{\i\omega\beta(\bx)}\bigl((\partial_3f_1^+(\bx,\bx_A,\omega))P^{-,+}(\bx,\bx_B,\omega)+
(\partial_3f_1^-(\bx,\bx_A,\omega))P^{+,+}(\bx,\bx_B,\omega)\bigr){\rm d}\bxh\nonumber
\end{eqnarray}
and 
\begin{eqnarray}
&&P^{+,+}(\bx_A,\bx_B,\omega)-\chi_A(\bxB)\{f_1^+(\bx_B,\bx_A,\omega)\}^*\ss(\omega)\label{eqAA61-alt}\\
&&=\int_{\setdD_0}\frac{-2}{\i\omega\beta(\bx)}\bigl(\{\partial_3f_1^+(\bx,\bx_A,\omega)\}^*P^{+,+}(\bx,\bx_B,\omega)+\{\partial_3f_1^-(\bx,\bx_A,\omega)\}^*P^{-,+}(\bx,\bx_B,\omega)\bigr){\rm d}\bxh,\nonumber
\end{eqnarray}
where $\chi_A$ is the characteristic function of the domain $\setDA$. It is defined by equation (\ref{eqC3.2D}), with $\setD$ \rev{and $\setdDD$ replaced by $\setD_A$ and $\setdDA$, respectively.}
Next, we replace the source in state $B$ by one which emits waves in the negative $x_3$-direction, hence
$B_B^+(\bx,\omega)=0$, $B_B^-(\bx,\omega)=\delta(\bx-\bx_B)\ss(\omega)$
and $\p_B^\pm(\bx,\omega)=P^{\pm,-}(\bx,\bx_B,\omega)$. This gives
\begin{eqnarray}
&&P^{-,-}(\bx_A,\bx_B,\omega)+\chi_A(\bxB)f_1^+(\bx_B,\bx_A,\omega)\ss(\omega)\label{eqAA62-alt}\\
&&=\int_{\setdD_0}\frac{2}{\i\omega\beta(\bx)}\bigl((\partial_3f_1^+(\bx,\bx_A,\omega))P^{-,-}(\bx,\bx_B,\omega)+(\partial_3f_1^-(\bx,\bx_A,\omega))P^{+,-}(\bx,\bx_B,\omega)\bigr){\rm d}\bxh\nonumber
\end{eqnarray}
and
\begin{eqnarray}
&&P^{+,-}(\bx_A,\bx_B,\omega)-\chi_A(\bxB)\{f_1^-(\bx_B,\bx_A,\omega)\}^*\ss(\omega)\label{eqAA63-alt}\\
&&=\int_{\setdD_0}\frac{-2}{\i\omega\beta(\bx)}\bigl(\{\partial_3f_1^+(\bx,\bx_A,\omega)\}^*P^{+,-}(\bx,\bx_B,\omega)+\{\partial_3f_1^-(\bx,\bx_A,\omega)\}^*P^{-,-}(\bx,\bx_B,\omega)\bigr){\rm d}\bxh.\nonumber
\end{eqnarray}
Equations (\ref{eqAA60-alt}) $-$ (\ref{eqAA63-alt}) are single-sided representations for backward propagation of the one-way wave fields $P^{\pm,\pm}(\bx,\bx_B,\omega)$ 
from $\setdD_0$ to $\bxA$.
Similar results have been previously obtained
%$^{47,48}$ 
\citep{Wapenaar2017GP2, Neut2017JASA}, 
but without decomposition at $\bxB$. An advantage of these equations over equations (\ref{eq104a-altb}) $-$ (\ref{eq104c-altb}) is that the backward propagated fields 
$P^{\pm,\pm}(\bx_A,\bx_B,\omega)$ are expressed entirely in terms of integrals along the surface $\setdD_0$.

Single-sided representations containing the field-normalised focusing functions $f_1^\pm(\bx,\bx_A,\omega)$ find applications for example  in reflection imaging methods which
account for multiple scattering. In these methods, the focusing functions are 
retrieved from the reflection response at the surface $\setdD_0$, using the Marchenko method
%$^{18,49-51}$ 
\citep{Wapenaar2014GEO, Singh2017GEO2, Costa2018GJI, Wapenaar2019SE}.

We conclude this section by considering a special case.
Suppose that in state $B$ the parameters $\alpha$ and $\beta$  are the same as in state $A$ throughout space.
Then $P^{\pm,\pm}(\bx,\bxB,\omega)=G^{\pm,\pm}(\bx,\bxB,\omega)\ss(\omega)$
for all $\bx$. Moreover,  $P^{+,\pm}(\bx,\bxB,\omega)=0$ for $\bx$ at $\setdD_0$. 
Substituting this into representations (\ref{eqAA60-alt}) $-$ (\ref{eqAA63-alt}), 
summing the left- and right-hand sides of these representations separately, dividing both sides by $\ss(\omega)$  
and using equation (\ref{eq101}), we obtain
\begin{eqnarray}
&&G(\bx_A,\bx_B,\omega)+\chi_A(\bxB)2\i\Im\{f_1(\bx_B,\bx_A,\omega)\}\nonumber\\
&&\hspace{0.5cm}=\int_{\setdD_0}\frac{2}{\i\omega\beta(\bx)}\partial_3\bigl(f_1^+(\bx,\bx_A,\omega)-\{f_1^-(\bx,\bx_A,\omega)\}^*\bigr)G^-(\bx,\bx_B,\omega){\rm d}\bxh\label{eqAA64-alt}
\end{eqnarray}
(with $\Im$ denoting the imaginary part), where $G^-(\bx,\bxB,\omega)=G^{-,+}(\bx,\bxB,\omega)+G^{-,-}(\bx,\bxB,\omega)$.
Taking the real part of both sides gives 
\begin{eqnarray}
&&G_{\rm h}(\bx_A,\bx_B,\omega)=\Re\int_{\setdD_0}\frac{4}{\i\omega\beta(\bx)}\partial_3\bigl(f_1^+(\bx,\bx_A,\omega)-
\{f_1^-(\bx,\bx_A,\omega)\}^*\bigr)G^-(\bx,\bx_B,\omega){\rm d}\bx,\nonumber\\&&\label{eqAA65}
\end{eqnarray}
where $G_{\rm h}(\bx_A,\bx_B,\omega)$ is the homogeneous Green's function, defined in equation (\ref{eq100}).
Unlike in equation (\ref{eq104acbd-altb}), here the homogeneous Green's function is represented by a single integral along the surface $\setdD_0$, containing field normalised one-way focusing
and Green's functions.

\section{Conclusions}\label{sec6}
We have considered flux-normalised and field-normalised decomposition of scalar wave fields into coupled one-way wave fields. The operators for field-normalised decomposition
exhibit less symmetry than those for flux-normalised decomposition. Nevertheless, we have shown that reciprocity theorems can be derived for field-normalised one-way wave fields
in a similar way as those for flux-normalised one-way wave fields. An additional condition for the reciprocity theorems for field-normalised one-way wave fields
is that in one of the states the derivatives in the $x_3$-direction of the parameters $\alpha$ and $\beta$ vanish at the boundary of the considered domain. 
This condition is easily fulfilled when one of the states is a Green's function or a focusing function, for which the parameters $\alpha$ and $\beta$ can be freely chosen at and outside the
boundary of the domain.

We have used the reciprocity theorems for field-normalised one-way wave fields as a starting point for deriving
representation theorems for field-normalised one-way wave fields in a systematic way. We obtained representations for forward and for backward propagation of one-way wave fields.
These representations account for multiple scattering.  
Whereas the Kirchhoff-Helmholtz integrals for forward propagation can be easily transformed into single-sided representations, this transformation is less straightforward for 
the Kirchhoff-Helmholtz integrals for backward propagation. By replacing the Green's functions by focusing functions we obtained single-sided representations for backward propagation 
of field-normalised one-way wave fields. These representations are particularly useful to retrieve wave fields in the interior of a domain in situations 
where measurements can be carried out only at a single surface. An important application is reflection imaging, accounting for multiple scattering.

\section*{Data Availability}

No datasets have been used for this study.

\section*{Conflicts of Interest}

The author declares that he has no competing interests.

\section*{Acknowledgements}
\rev{I thank an anonymous reviewer for the constructive review, which helped to improve the paper.}
This work has received funding from the European Union's Horizon 2020 research and innovation programme: European Research Council (grant agreement 742703).

%\bibliography{/Domain/tudelft.net/Users/cwapenaar/kees/lib/tex/bibliography/bibliography}

%\end{spacing}

%\end{document}

\end{spacing}

\end{document}